\documentclass[twocolumn]{aastex61}

\usepackage{epsfig}
\usepackage{amsmath}
\usepackage{amssymb}
\usepackage{graphicx}
\usepackage{enumerate}
\usepackage{placeins}
\usepackage{verbatim}

\newcommand{\archain}{AR-CHAIN}
\newcommand{\ketju}{KETJU}
\newcommand{\gadget}{GADGET-3}

\newcommand{\abs}[1]{\left|#1\right|}
\newcommand{\vect}[1]{\boldsymbol{#1}}
\newcommand{\bigO}{\mathcal{O}} 

\newcommand{\ud}{\mathrm{d}}
\newcommand{\derfrac}[2]{\frac{\ud #1}{\ud #2}}

\defcitealias{Thomas2016}{T16}
\defcitealias{Rantala2017}{R17}

\received{DD.MM.YYYY}
\revised{DD.MM.YYYY}
\accepted{DD.MM.YYYY}
\submitjournal{ApJ}

\shorttitle{Core scouring by supermassive black holes}
\shortauthors{Rantala et al.}
\begin{document}
\title{The formation of extremely diffuse galaxy cores\\ by merging 
supermassive 
black 
holes}

\correspondingauthor{Antti Rantala}
\email{antti.rantala@helsinki.fi}

\author[0000-0001-8789-2571]{Antti Rantala}
\affil{Department of Physics, University of Helsinki, Gustaf H\"allstr\"omin 
katu 2a, 00560 Helsinki, Finland}

\author{Peter H. Johansson}
\affiliation{Department of Physics, University of Helsinki, Gustaf 
H\"allstr\"omin katu 2a, 00560 Helsinki, Finland}

\author{Thorsten Naab}
\affiliation{Max-Planck-Insitut f\"ur Astrophysik, Karl-Schwarzchild-Str. 1, 
D-85748, Garching, Germany}

\author{Jens Thomas}
\affiliation{Max-Planck-Institut f\"ur Extraterrestriche Physik, 
Giessenbach-Str. 1, D-85741, Garching, Germany}

\author{Matteo Frigo}
\affiliation{Max-Planck-Insitut f\"ur Astrophysik, Karl-Schwarzchild-Str. 1, 
D-85748, Garching, Germany}

\begin{abstract}

Given its velocity dispersion, the early-type galaxy NGC 1600 has an unusually 
massive ($M_\bullet = 1.7 \times 10^{10} M_\odot$) central supermassive black 
hole (SMBH), surrounded by a large core ($r_\mathrm{b} = 0.7$ kpc) with a 
tangentially biased stellar 
distribution. We present high-resolution equal-mass merger simulations 
including 
SMBHs to study the formation of such systems. The structural parameters of the 
progenitor ellipticals were chosen to produce merger remnants resembling NGC 
1600. We test initial stellar density slopes of $\rho 
\propto r^{-1}$ and $\rho \propto r^{-3/2}$ and vary the initial SMBH masses 
from $8.5 \times 10^8$ to $8.5 \times 10^9$ $M_\odot$. With increasing SMBH 
mass 
the 
merger remnants show a systematic decrease in central surface brightness, an 
increasing core size, and an increasingly tangentially biased central velocity 
anisotropy. Two-dimensional kinematic maps reveal decoupled, rotating core 
regions for the most massive SMBHs. The stellar cores form rapidly as the SMBHs 
become bound, while the velocity anisotropy develops more slowly after the SMBH 
binaries become hard. The simulated merger remnants follow distinct relations
between the core radius and the sphere-of-influence, and the SMBH mass, 
similar to observed systems. We find a systematic change in the relations 
as a function of the progenitor density slope, and present a simple scouring 
model reproducing this behavior. Finally, we find the best agreement with NGC 
1600 using SMBH masses totaling the observed value of $M_\bullet = 1.7 \times 
10^{10} M_\odot$. In general, density slopes of $\rho \propto r^{-3/2}$ for the 
progenitor galaxies are strongly favored for the equal-mass merger scenario.  

\end{abstract}


\section{Introduction} \label{section: intro}

Observations have revealed a dichotomy in the population of bright elliptical 
galaxies, with the 
brighter ellipticals $(M_{B} \lesssim -20.5)$ being dominated by systems with 
little rotation (slow-rotators), 
typically boxy isophotes and relatively shallow central surface brightness 
profiles, whereas 
intermediate luminosity ellipticals $(-20.5 \lesssim  M_{B} \lesssim -18.5)$ 
have more rotational support (fast-rotators), 
more disky isophotes and steeper power-law like central surface brightness 
profiles (e.g. \citealt{Kormendy1996,Faber1997}). 

These morphological differences are also indicative of two distinct formation 
paths for early-type galaxies. 
Lower luminosity elliptical galaxies likely formed through in-situ star 
formation and mergers of gas-rich disk-dominated galaxies, with the 
accompanying 
merger induced star-burst resulting in
cuspy central stellar profiles (e.g. 
\citealt{Barnes1996,Naab2006,Cappellari2007,Hopkins2008,Johansson2009,
Krajnovic2011,Cappellari2016,Lahen2018}). 
The more massive early-type galaxies are instead believed to have assembled 
through 
a 
two-stage process, in which the early  assembly is dominated by rapid in situ 
star formation fueled by cold gas flows 
and hierarchical merging of multiple star-bursting progenitors, whereas the 
later growth below redshifts of $z\lesssim 2-3$ is dominated by a more 
quiescent 
phase
of gas-poor (dry) merging, in which the galaxy accretes stars formed mainly in 
progenitors outside the main galaxy 
(e.g. 
\citealt{Naab2009,oser2010,2011Feldmann,Johansson2009c,2012Johansson,Moster2013,
2015Wellons,rodriguez-gomez2016}, see also \citealt{Naab2017} for a review). 

In general the smooth light profiles of elliptical galaxies over several orders 
of magnitude in radius 
are well described by a single three-parameter S\'ersic profile 
\citep{Sersic1963, Caon1993}. However, the most massive
ellipticals often exhibit in their central light profiles significantly flatter 
regions of missing light, when compared to the inwards extrapolation of 
the  outer S\'ersic profile. The radius at which this departure from the 
S\'ersic 
profile occurs is often called the ''break'' or ''core'' radius, denoted by 
$r_\mathrm{b}$.    
Typically the size of the core region is $r_\mathrm{b}\sim 50 -500 \ \rm pc$ 
(e.g. \citealt{Ravindranath2002,Lauer2007,Rusli2013,Dullo2014}), but in extreme cases the 
core can 
extend beyond the kpc-scale 
(e.g. \citealt{Postman2012, LopezCruz2014,Dullo2017}). 

At face value the existence of cores in massive galaxies poses a theoretical 
challenge to their formation scenario. The reason for this is that the
central structure of a merger remnant without gas is dominated by the more 
concentrated of the two progenitors, meaning that the steeper central density 
cusp survives 
(e.g. \citealt{Holley-Bockelmann1999,Boylan-Kolchin2004}). 
The massive ellipticals are believed to have been assembled from mergers of 
lower-mass ellipticals with steep central-density cusps at high redshifts. 
Situations in which the merger progenitors contain gas complicate the problem, 
as the angular momentum transfer in a merger leads to gas inflows that trigger 
central starbursts, resulting in the formation of even denser central regions 
(e.g. 
\citealt{Barnes1991}). Thus 
the shallow cores widely observed in massive early-type galaxies must result 
from another physical process than galaxy merging. 

The supermassive black holes (SMBHs) found in the centers of all massive 
galaxies in 
the 
local Universe (e.g. \citealt{Kormendy1995,Ferrarese2005,kormendy2013}) could 
potentially provide 
a core formation mechanism. In particular, there is evidence for a 
co-evolution of SMBHs and their host galaxies 
as manifested in the surprisingly tight relations between the SMBH masses and 
the fundamental properties of the galactic bulges that host them, e.g. the 
bulge 
mass (e.g. \citealt{Haring2004}) and the bulge stellar velocity dispersion 
(e.g. \citealt{Ferrarese2000,Gebhardt2000,Tremaine2002}). In addition there are also 
indications for a core-black hole mass relation as the size of the core 
and the amount of ''missing'' 
starlight in the center of the core galaxy approximately scale with the mass of 
the central black hole \citep{Graham2004,deRuiter2005,Lauer2007,Kormendy2009,Rusli2013,Dullo2014,Thomas2016}.   

The formation process of cores in massive elliptical galaxies is commonly 
attributed to core scouring by SMBH binaries in the aftermath of galaxy 
mergers. 
After sinking to 
the center of the merger remnant by dynamical friction 
\citep{Chandrasekhar1943} 
from surrounding stars and dark matter, the two SMBHs form a 
gravitationally bound binary. The binary subsequently shrinks by interacting 
with the stellar background and ejecting stars in complex three-body 
interactions, which 
carry away energy and angular momentum from the SMBH binary system (e.g. 
\citealt{hills1980}). The core scouring process also affects the orbit 
distribution of the stars, as 
only stars on radial orbits that get sufficiently close to the SMBH binary can 
be ejected. Consequently, the orbital structure in the core after the scouring 
process is expected 
to be strongly biased towards tangential orbits, with the ejected stars 
contributing to enhanced radial motions outside the core region 
\citep{Quinlan1997,Milosavljevic2001}.
Finally, if the so-called final-parsec problem is avoided, which describes the 
depletion of the center-crossing (or 'SMBH loss cone') orbital population, 
gravitational 
wave emission becomes important and the binary radiates the remaining orbital 
energy and angular momentum, merging into 
a single SMBH in a burst of gravitational radiation (e.g. 
\citealt{Begelman1980, 
Ebisuzaki1991, Milosavljevic2001, Milosavljevic2003, Volonteri2003, Merritt2005, Merritt2013}).

In this paper we study the formation of cores in 
collisionless gas-poor (dry) numerical merger simulations. Technically this is 
a 
demanding problem as 
we need to simultaneously follow the global galactic-scale dynamical processes, 
while solving accurately the dynamics of SMBHs, SMBH binaries, and the 
surrounding stellar systems 
down to subparsec scales. Traditionally these types of problems have been 
studied 
using N-body codes that calculate the gravitational force directly by 
summing the force from every particle on every particle (e.g. 
\citealt{Aarseth1999}). Additionally, the close encounters of simulation 
particles are usually treated in few-body subsystems with very high accuracy 
(e.g. 
\citealt{Aarseth2003}). The advantages of this approach is high numerical 
accuracy and the fact that       
this method does not require gravitational softening unlike smoothed particle 
hydrodynamics (SPH) tree codes (e.g. 
\citealt{SpringelDiMatteo2005,Johansson2009b}) and 
adaptive mesh refinement codes (e.g. \citealt{Kim2011,Sijacki2015}) which
have been commonly used to study black hole dynamics in global scale 
simulations. A significant drawback of the direct N-body approach is the 
required computational time, which
scales steeply with the particle number $\bigO(N^2)$ as opposed to tree and 
mesh 
codes, which typically scale as $\bigO(N \log N)$. Thus, numerical simulations 
with N-body codes have 
typically only been able to explore separate aspects of the full problem by 
limiting themselves to studies of SMBH binary dynamics in the centers
of isolated galaxies or merger remnants, with the surrounding galaxy often 
represented by idealized initial conditions (e.g. 
\citealt{Milosavljevic2001,Milosavljevic2003,berczik2006,Preto2011,khan2011,
khan2013,Gualandris2012,Vasiliev2014b,Holley_Bockelmann2015,Wang2016}).

All the simulations in this paper are run with our recently developed hybrid 
tree-N-body code \ketju{} (the word for 'chain' in 
Finnish) (\citealt{Rantala2017}, hereafter \citetalias{Rantala2017}).
This code combines an algorithmic chain regularization (\archain) 
\citep{Mikkola2006,Mikkola2008}
method to efficiently and accurately compute the dynamics close to SMBHs with 
the fast and widely used tree code \gadget \  \citep{Springel2005} for the 
calculation of the global galactic dynamics.

The properties of our initial conditions (ICs) have been motivated by the 
recent 
observations and dynamical
modeling of the core galaxy NGC 1600 presented in \citet{Thomas2016} (hereafter 
\citetalias{Thomas2016}) as part of the MASSIVE survey \citep{Ma2014}. NGC 1600 
is a relatively isolated elliptical galaxy 
near the
center of a galaxy group at a distance of $64 \ \rm Mpc$, (assuming $H_0 = 73 \ 
\rm km/s/Mpc)$ and has a stellar mass of $M_\star=8.3 \times 10^{11} M_{\odot}$ 
and dark matter halo mass of $M_\mathrm{DM} = 1.5 \times 10^{14}$ $M_\odot$. 
This galaxy shows little rotation, with $v_{\rm rot} < 30 \ \rm km/s$ and the 
line-of-sight velocity dispersion rises from $\sigma=235 - 275 \ \rm km/s$ at 
large 
radii to $\sigma=359 \ \rm km/s$ near the very center (\citealt{Bender1994}; 
\citetalias{Thomas2016}). NGC 1600 is most likely the result of a 
collisionless, 
low-angular momentum binary merger, which took place several gigayears ago as 
indicated by the low level of present star formation and the dynamical state of
the galaxy \citep{Matthias1999,Smith2008}. What makes NGC 1600 very interesting 
is the remarkably large, faint and flat core 
(e.g. \citealt{Lauer1985, Quillen2000, Lauer2007}, \citetalias{Thomas2016}) of size $r_\mathrm{b}\sim 
0.7 \ \rm kpc$. In addition the recent dynamical modeling by 
\citetalias{Thomas2016}
determined the mass of the central supermassive black hole to be very large at 
$M_\bullet \sim 1.7 \times 10^{10} M_\odot$ constituting 2.1\% of the total 
stellar 
bulge mass, well above the 0.1\% nominally expected from the 
$M_{\bullet}-M_{\rm 
bulge}$ relation.  

For simplicity, we focus in this study on a single generation of binary galaxy 
mergers ($N_\mathrm{mergers}=1$). The more realistic (and computationally 
far more demanding) scenario $N_\mathrm{mergers}>1$ is left for future work. 
Earlier studies on the stellar mass deficit $M_\mathrm{def}$ displaced by 
merging 
SMBH binaries indicate $N_\mathrm{mergers} \gtrsim 2$, as N-body considerations 
suggest $M_\mathrm{def} \sim 0.5 \times N_\mathrm{mergers} 
\times 
M_\bullet$ 
(e.g. \citealt{Merritt2006}) while observations (e.g. \citealt{Kormendy2009, 
Rusli2013,Dullo2014}) point towards $M_\mathrm{def} \sim 1 - 10 \times M_\bullet$. 
We stress that the comparison of our simulation results to the actual 
observations should be done keeping the assumption of a single merger 
generation 
in mind.

We begin this article by briefly reviewing the main features of the \ketju{} 
simulation code in \S \ref{section: ketju}. In the following \S \ref{section: 
simulations} we describe our initial conditions and discuss the 
merger sample. The main results concerning the core scouring process is 
discussed in \S \ref{section: corescouring}. We also compare our 
simulated surface brightness and velocity anisotropy profiles with the 
corresponding
observed profiles for NGC 1600. In \S \ref{section:scaling_relations} we 
discuss the origin of the tight scaling relations between the observed 
galaxy core sizes $r_\mathrm{b}$, the SMBH masses $M_\bullet$ and the spheres 
of 
influence of the SMBHs $r_\mathrm{SOI}$, first presented in 
\citetalias{Thomas2016}.
Finally, we present our conclusions in \S \ref{section: conclusions}.

\section{Numerical code}\label{section: ketju}
\subsection{The \ketju{}  code}
\label{section: ketju_code}

Our simulations are run using the recently developed \ketju{} code, which  
is built on the widely-used galaxy simulation code \gadget{} 
\citep{Springel2005}.
Here we briefly review the main features of the code and refer the reader to 
\citetalias{Rantala2017} for a complete description of the code.
The central idea of the \ketju{} code is to include special regions around 
every 
black hole particle, in which the dynamics of SMBHs and stellar particles is 
modeled using
a non-softened algorithmic chain regularization technique \citep{Mikkola2006, 
Mikkola2008} that also includes Post-Newtonian (PN) 
corrections (e.g. \citealt{Will2006}). The remaining particles far from the 
SMBHs are evolved 
using the standard \gadget{} leapfrog integrator, with the softened 
gravitational 
force calculated either using  
a pure tree or a hybrid tree-mesh TreePM algorithm (\citealt{Barnes1986}; 
\citealt{Springel2001}). 

The \ketju{} code operates by dividing the simulation particles into three 
categories (see Fig. 2 in \citetalias{Rantala2017} for 
an illustration). All the SMBH particles and the stellar particles, which lie 
within a user defined chain radius $(r_{\rm chain})$ are marked 
as chain subsystem particles. Particles that lie just outside the chain radius, 
but 
induce a strong tidal perturbation on the chain system 
are marked as perturber particles. Finally, the remaining particles that are 
far from any SMBHs are treated as ordinary \gadget{} particles 
with respect to the force calculation. The chain membership of the simulation 
particles is evaluated at the beginning of each timestep. \ketju{} also allows 
for both multiple simultaneous chain subsystems and several SMBHs in a single 
subsystem.

The chain radius should be chosen carefully to ensure that the regularized 
regions are large enough to accurately simulate
the dynamics around SMBHs, but at the same time 
small enough in order not to excessively degrade the performance of the code. 
The minimum chain radius is set by the Plummer-equivalent gravitational 
softening length in \gadget:
$r_{\mathrm{chain}} > 2.8 \times \epsilon$, which ensures that the star-SMBH 
and 
SMBH-SMBH interactions always remain non-softened in \ketju. 
The maximum chain radius is limited by the number of particles in the chain 
subsystem. The scaling of the \archain{} integrator 
with particle number is of the order of $\bigO{(N^{2.13-2.20})}$ 
\citepalias{Rantala2017}. In the current implementation the chain region is 
limited 
to roughly $N_\mathrm{c} \sim 500$ particles for stellar density profiles 
$\rho_\star \propto r^{-1}$, beyond which the simulations 
becomes unfeasible to run. Thus for steeper density profiles 
$\rho_\star \propto r^{-\gamma}$, with $\gamma>1$, and higher numerical 
resolution the chain radius must 
be reduced in size. For typical applications presented in this paper 
the chain radius, $r_{\mathrm{chain}}$ is set at a few tens of parsecs, 
resulting in a few hundred particles within a single chain region. 

The regularized subsystems are also tidally perturbed by their surroundings, 
imposing a perturbing acceleration $\vect{f}_i$ on every 
particle $i$ in the subsystem. Particles in the immediate vicinity of a chain 
subsystem are predicted forward in 
time during the \archain{} external perturbations calculation, while the tidal 
field 
from the more distant particles remains constant during the \archain{} 
integration. 
The distinction between the nearby perturbing particles and the 
particles in the far-field in set by the perturber radius
\begin{equation}
	r_{\mathrm{pert}} = \gamma_{\mathrm{pert}} \times \left( 
\frac{m}{M_\bullet} 
\right)^{1/3} \times r_{\mathrm{chain}},
\end{equation}
in which $m$ is the mass of the perturbing simulation particle and $M_\bullet$ 
is the mass of the SMBH in the chain subsystem. The constant 
$\gamma_{\mathrm{pert}}$ is typically chosen so that $r_{\mathrm{pert}} = 2 
\times r_{\mathrm{chain}}$.

\subsection{The regularized integrator}

During each global \gadget{} timestep the particles in the chain subsystems are 
propagated using a novel re-implementation of the \archain{} algorithm 
\citep{Mikkola2008} 
by \citetalias{Rantala2017}.
The \archain{} algorithm has the following three main aspects. Firstly, the 
equations of motion of the simulation particles are time-transformed using a 
new 
time coordinate 
in the regularized regions (e.g. \citealt{Mikkola2002}), with the integration 
proceeding using a leapfrog algorithm. In essence, algorithmic regularization 
works by transforming 
the equations of motion by introducing a fictitious time variable such that 
integration by the common leapfrog method yields 
exact orbits for a Newtonian two-body problem including two-body collisions. 
Secondly, numerical round-off errors are greatly reduced by the introduction of 
a coordinate system based on chained inter-particle 
vectors (e.g. \citealt{Mikkola1993}). Finally, the Gragg-Bulirsch-Stoer 
\citep{Gragg1965,Bulirsch1966} extrapolation method yields 
high numerical accuracy in orbit integrations at a preset user-given error 
tolerance level 
$(\eta_\mathrm{GBS})$.  
 
The regularized integrator has multiple advantages compared to the standard 
\gadget{} leapfrog algorithm, when integrating the equations of motion of the 
simulation particles near SMBHs. In addition to high numerical accuracy and 
sophisticated 
error control, gravitational softening is not applied in the regularized 
regions. 
Thus, the N-body dynamics 
is accurately resolved even at very small particle separations. Furthermore, 
the extension of phase space with 
an auxiliary velocity variable in the algorithm (\citealt{Hellstrom2010}, 
\citealt{Pihajoki2015}) allows for 
the efficient implementation of the velocity-dependent Post-Newtonian 
corrections in the  motion of the simulation particles (e.g. 
\citealt{Will2006}). In this study, we include PN terms for the SMBH-SMBH 
interaction only, since the star-SMBH terms
would be unphysically large, given that the stellar particles have very large 
masses, $m_\star \gg M_\odot$. In the included PN terms we go up to
order PN2.5, the highest of which includes the radiation reaction term due to 
losses 
caused by the emission of gravitational waves.  
Higher-order PN corrections, spin terms and their cross terms are not used in 
this study.

To summarize, \ketju{} is able to follow SMBH dynamics accurately from global 
galactic scales to  
particle separations of a few tens of Schwarzschild radii of the SMBHs where 
the 
PN approach breaks down. Before this happens
the SMBH particles are merged using a merger criterion that is based on the 
analytic \citet{Peters1963} estimate for the 
gravitational wave driven decay of the orbital binary semi-major axis (see Eq. 
\ref{eq: peters}). 
For additional details concerning 
the \archain{} integrator, see Appendices A and B in \citetalias{Rantala2017}.

\subsection{\ketju{} code updates}

Here we present code updates compared to the previous code version 
presented in 
\citetalias{Rantala2017}. The far-field acceleration part of the tidal 
acceleration 
$\vect{f}_i$ is now calculated individually for every particle in the 
regularized subsystem. In the previous version, the far-field acceleration was 
imposed on the center-of-mass of the chain subsystem only. 

In addition, the regularized subsystems 
are now temporarily deconstructed for the \gadget{} tree gravity computation. 
The gravitational forces between particles in the same subsystem are ignored in 
the tree calculation, and the 
center-of-mass of the regularized subsystem is propagated as an ordinary 
\gadget{} SMBH particle in the simulation. This procedure eliminates the demand 
for a distinct force correction used in \cite{Karl2015} and the previous 
version of the \ketju{} code. The old force correction procedure included a 
step in which the tree force on the center-of-mass of a subsystem was canceled 
using 
a directly summed term. This step produced low-level spurious noise to the 
total 
gravitational force as the directly summed component and the tree component did 
not always cancel exactly. Overall, the code updates have only a minor effect 
on the simulation results but are added for the sake of code clarity, and to 
ensure a consistent formulation of Newtonian gravity that also results in a 
slight performance upgrade.

Most importantly, the interface between \gadget{} and the \archain{} integrator 
is rewritten for an increased performance for simulations with high particle 
numbers, in excess of $N \gtrsim 5 \times 10^6$. In addition, the 
regularized subsystems are now built and deconstructed every smallest \gadget{} 
timestep. Now the tree rebuild on every smallest \gadget{} timestep is no 
longer 
necessary. Nevertheless, the tree domain update frequency is kept smaller than 
the typical value $\sim 0.05$ in \gadget{} \citep{Springel2005}. In this 
study, we use the value of $0.01$. This code interface formulation anticipates 
the future use of \ketju{} for hydrodynamical 
simulations that also include \gadget{} sub-resolution feedback modeling 
\citep{Hu2014, 
Hu2016}.
However, in the present study we still restrict ourselves to simulating 
collisionless 
gas-poor (dry) galaxy mergers that include central SMBHs.

\section{Numerical simulations}\label{section: simulations}
\subsection{Multi-component initial conditions}\label{section: ic}

In setting up the galaxies we use the Dehnen density-potential 
\citep{Dehnen1993}
models, which are commonly used for modeling the stellar components of 
early-type galaxies and bulges.
The spherically symmetric one-component Dehnen model is 
defined by three parameters: the total mass $M$, the scale radius $a$ and the 
central slope of the density profile $\gamma$. The allowed values for the 
central slopes $\gamma$ span $0 \leq \gamma < 3$. The most commonly used Dehnen 
models are the Hernquist profile with $\gamma = 1$ \citep{Hernquist1990}, the 
Jaffe profile with $\gamma = 2$ \citep{Jaffe1983} and the $\gamma = 3/2$ 
profile, which closely resembles the de Vaucouleurs profile 
\citep{deVaucouleurs1948, Dehnen1993} when projected. 

The Dehnen density-potential pair family is defined as 
\begin{equation}
\rho(r) = \frac{(3-\gamma) M}{4 \pi} \frac{a}{r^\gamma (r+a)^{4-\gamma}}
\label{eq: dehnen-density}
\end{equation}
\begin{equation}
\phi(r) =\frac{G M}{a} \times \begin{cases}
-  \frac{1}{2-\gamma} \left[ 1 - \left( \frac{r}{r+a} \right)^{2-\gamma} 
\right], & \gamma \neq 2\\
\ln {\frac{r}{r+a}}, & \gamma = 2.
\end{cases}
\label{eq: dehnen-potential}
\end{equation}
From the density profile we can solve the cumulative mass profile $M(r)$ 
and the three-dimensional half-mass radius $r_{1/2}$ 
\begin{equation}
 M(r) = 4 \pi \int_0^r \rho(r) r^2 dr = M \left( \frac{r}{r+a} 
\right)^{3-\gamma}
\label{eq: dehnen-mass}
\end{equation}
\begin{equation}
 r_{1/2} = a \left(2^{1/(3-\gamma)} -1 \right)^{-1},
\label{eq: dehnen-halfmass}
\end{equation} 
whereas the projected half-mass (effective) radius $R_{\mathrm{e}}$ can be well 
approximated by $R_{\mathrm{e}} \approx 3/4$ $r_{1/2}$.

We construct spherically symmetric, isotropic multi-component initial 
conditions (IC) consisting of a stellar bulge, a dark matter (DM) halo and a 
central SMBH. 
The initial conditions are generated using the distribution function method 
(e.g. \citealt{Merritt1985, Ciotti1992}) following the approach of 
\citet{Hilz2012}.
The distribution functions $f_i$ 
for the different components of the galaxy are obtained from the corresponding 
density profile 
$\rho_i$ and the total gravitational potential $\Phi_{\mathrm{T}}$ 
using Eddington's formula \citep{Binney2008}:
\begin{equation}
	f_i \left(\mathcal{E} \right) = \frac{1}{\sqrt{8} \pi^2} 
\int_{\Phi_{\mathrm{T}} = 0}^{\Phi_{\mathrm{T}} = \mathcal{E}} \frac{d^2 
\rho_i}{d \Phi_{\mathrm{T}}^2} \frac{d 
\Phi_{\mathrm{T}}}{\sqrt{\mathcal{E}-\Phi_{\mathrm{T}}}},
\label{eq: eddington}
\end{equation}
in which $\mathcal{E} = -\frac{1}{2} v^2 - \Phi_{\mathrm{T}} + \Phi_0$ is the 
(positive) energy relative to the chosen zero point of the potential $\Phi_0$. 
In general, the zero point is chosen so that $f_i > 0$ for $\mathcal{E} > 0$ 
and $f_i = 0$ for $\mathcal{E} \le 0$. For an isolated system extending to 
infinity, such as the Dehnen profiles in our case, we can set $\Phi_0 = 0$. 
The particle positions of the components are drawn from the cumulative mass 
profile (Eq. \ref{eq: dehnen-mass}), after which the velocities of the 
particles are sampled from pre-computed and tabulated distribution functions 
\citepalias{Rantala2017}.

The SMBH is represented by a single point mass $M_\bullet$ at rest at the 
origin. The 
stellar component is parameterized by the total stellar mass $M_\star$, the 
effective radius of the stellar component $R_{\mathrm{e}}$ and the central 
density profile slope $\gamma$. The dark matter halo is modeled as a Hernquist 
sphere 
($\gamma = 1$) with a total mass $M_{\mathrm{DM}}$ 
\citep{SpringelDiMatteo2005}. 
The scale radius of the DM component is computed assuming a dark matter 
fraction $f_\mathrm{DM}$ inside the three-dimensional stellar half-mass radius 
$r_{1/2}$. Defining the DM fraction inside a radius $r$ as
\begin{equation}
f_\mathrm{DM}(r) = \frac{M_\mathrm{DM}(r)}{M_\star(r) + M_\mathrm{DM}(r)}
\label{eq: dmfrac}
\end{equation}
the scale radius of the DM component can be derived by inserting Eq. \eqref{eq: 
dehnen-mass}
describing the cumulative mass profile into  Eq. \eqref{eq: dmfrac} and using 
the definition of the half-mass 
radius from Eq. \eqref{eq: dehnen-halfmass}, resulting in
\begin{equation}
a_\mathrm{DM} \approx \frac{4}{3} \left[ \sqrt{\frac{2 M_\mathrm{DM}}{M_\star} 
\left( \frac{1}{f_\mathrm{DM}(r_{1/2})} -1 \right) } -1 \right] R_{\mathrm{e}}.
\end{equation}

\subsection{Progenitor galaxies and merger orbits}

We study a simplified scenario in which the current structural properties of 
massive core elliptical galaxies were shaped in a final dry major merger of two 
early-type galaxies. The progenitor galaxies are identical to each other in 
each 
simulation 
run. The structural properties of the progenitors are motivated by the 
observations 
and dynamical 
modeling of the core elliptical galaxy NGC 1600 \citepalias{Thomas2016}. In 
Table \ref{table: progenitors} we list the physical parameters of the 
progenitor 
models ($M_\star$, $R_\mathrm{e}$, $M_\mathrm{DM}$, $f_\mathrm{DM}$), which 
remain the 
same for every 
IC in this study. We assume that all the stellar mass was already present in 
the progenitor galaxies at the time of this final major merger, yielding 
progenitor stellar and dark matter halo masses of $M_\star = 4.15 \times 
10^{11}$ $M_\odot$ 
and
$M_\mathrm{DM} = 7.5 \times 10^{13}$ $M_\odot$ \citepalias{Thomas2016}. 

Using virial arguments \citet{Naab2009} showed that the effective radius of a 
galaxy grows 
in a dry equal-mass major by roughly a factor of $\sim 2$. We set the effective 
radius of 
our progenitor galaxies to $R_\mathrm{e} = 7$ kpc, as observations have shown 
that the effective radius 
of NGC 1600 is $R_{\mathrm{e}} \sim14$-$16$ kpc (e.g. \citealt{Matthias1999, 
Trager2000, Fukazawa2006}).
The dark matter fraction was set to $f_\mathrm{DM} = 0.25$, which is a rather 
conservative estimate, as the observed 
central DM fractions inside $R_\mathrm{e}$ of local early-type galaxies are 
typically 
found to be 
relatively low (e.g. \citealt{Cappellari2013, Courteau2015}).
\begin{table}
\begin{center}
\begin{tabular}{ | c | c | c |}
    \hline
    Parameter & Symbol & Value\\ \hline
    Stellar mass&$M_\star$               & $4.15 \times 10^{11}$ $M_\odot$\\ 
    Effective radius&$R_\mathrm{e}$     & $7$ kpc\\ 
    DM halo mass&$M_\mathrm{DM}$  & $7.5 \times 10^{13}$ $M_\odot$\\ 
    DM fraction&$f_\mathrm{DM}(r_{1/2})$ & 0.25\\  
    Number of stellar particles&$N_\star$ & $4.15 \times 10^6$\\ 
    Number of DM particles&$N_\mathrm{DM}$ & $1.0 \times 10^7$\\
    \hline
\end{tabular}
\caption{The physical properties of the merger progenitor galaxies used for all 
simulations in this study. 
The definitions of the parameters are described in the main text.}
\label{table: progenitors}
\end{center}
\end{table}

We construct a sample of 14 initial conditions for major merger 
simulations. In half of the sample the stellar component of the progenitor 
galaxies is 
modeled using a Hernquist profile  $(\gamma = 1)$ and in the other half of ICs 
we use a steeper $\gamma = 
3/2$ profile. All the 
progenitor models are listed in Table \ref{table: ics}. 

For 12 initial conditions we include an SMBH with initial masses ranging from 
$M_\bullet = 8.5 \times 10^{8}$ $M_\odot$ to 
$M_\bullet = 8.5 \times 10^{9}$ $M_\odot$.
For comparison, we also included one set of models without a central SMBH. 
The largest SMBH masses are consistent with the final SMBH mass obtained using 
the dynamical modeling of NGC 1600 ($M_\bullet =1.7 \times 10^{10}$ $M_\odot$) 
while the ICs with lower-mass SMBHs lie closer 
to the observed $M_\bullet - \sigma$ -relation \citepalias{Thomas2016}. Note 
that even if the 
stellar and the DM density profiles are identical in the ICs with different 
SMBH masses, the 
circular velocity and velocity dispersion profiles differ due to the different 
gravitational potential of the SMBHs, especially near the central regions of 
the galaxies.
\begin{table}
\begin{center}
\begin{tabular}{ | c | c | c | }
    \hline
    Progenitor & $\gamma$ & $ M_\bullet $\\ \hline
    $\gamma$-1.0-BH-0 & 1.0 & -\\
    $\gamma$-1.0-BH-1 & 1.0 & $8.5 \times 10^{8}$ $M_\odot$\\ 
    $\gamma$-1.0-BH-2 & 1.0 & $1.7 \times 10^{9}$ $M_\odot$\\ 
    $\gamma$-1.0-BH-3 & 1.0 & $3.4 \times 10^{9}$ $M_\odot$\\ 
    $\gamma$-1.0-BH-4 & 1.0 & $5.1 \times 10^{9}$ $M_\odot$\\ 
    $\gamma$-1.0-BH-5 & 1.0 & $6.8 \times 10^{9}$ $M_\odot$\\ 
    $\gamma$-1.0-BH-6 & 1.0 & $8.5 \times 10^{9}$ $M_\odot$\\
    $\gamma$-1.5-BH-0 & 1.5 & -\\
    $\gamma$-1.5-BH-1 & 1.5 & $8.5 \times 10^{8}$ $M_\odot$\\ 
    $\gamma$-1.5-BH-2 & 1.5 & $1.7 \times 10^{9}$ $M_\odot$\\ 
    $\gamma$-1.5-BH-3 & 1.5 & $3.4 \times 10^{9}$ $M_\odot$\\ 
    $\gamma$-1.5-BH-4 & 1.5 & $5.1 \times 10^{9}$ $M_\odot$\\ 
    $\gamma$-1.5-BH-5 & 1.5 & $6.8 \times 10^{9}$ $M_\odot$\\ 
    $\gamma$-1.5-BH-6 & 1.5 & $8.5 \times 10^{9}$ $M_\odot$\\
    \hline
\end{tabular}
\caption{The Dehnen model $\gamma$-coefficients used for modeling the stellar 
profiles and 
the initial SMBH masses of the 14 progenitor galaxies used in this study.}
\label{table: ics}
\end{center}
\end{table}

Similarly to \citetalias{Rantala2017}, the progenitor galaxies are set on a 
nearly-parabolic 
merger orbit with an initial separation of $d=30$ kpc. From this orbit, the 
approach of the galaxies is swift and the central stellar cusps merge before $t 
\sim 300$ Myr.

\subsection{Accuracy parameters and numerical resolution}

The values of the \gadget{} and \archain{} numerical accuracy parameters are 
based on 
the simulations of \citetalias{Rantala2017}.
We set the \gadget{}  integrator error tolerance to $\eta = 0.002$ and the 
force accuracy to $\alpha = 0.005$, using the standard cell opening criterion 
\citep{Springel2005}. For the \archain{} integrator, we choose a 
Gragg-Bulirsch-Stoer (GBS) tolerance of $\eta_\mathrm{GBS}= 10^{-6}$. 

The chain radius is set to $r_\mathrm{chain} = 30$ pc in the runs 
with an initial slope for the stellar density profile of $\gamma = 1$ and 
$r_\mathrm{chain} = 10$ pc for the runs with a steeper $\gamma = 3/2$ slope. 
These choices for 
the chain radii result in a few hundred particles in the regularized regions 
at the start of the simulations, making the runs numerically feasible, as 
discussed in
\S \ref{section: ketju_code}. The gravitational softening lengths for the 
\gadget{} stellar particles are 
selected accordingly: $\epsilon_\star = 10$ pc in the $\gamma = 1$ runs and 
$\epsilon_\star = 3.5$ pc in the $\gamma = 3/2$ 
simulations, chosen to fulfill the criterion $r_\mathrm{chain} > 2.8 \times 
\epsilon$. The 
softening length for the dark matter component is $\epsilon_\mathrm{DM} = 100$ 
pc in all 
the simulation runs. The tidal 
perturber parameter $\gamma_{\mathrm{pert}}$ is set following the condition 
described in \S \ref{section: ketju_code}, resulting in $r_{\mathrm{pert}} 
= 2 \times r_{\mathrm{chain}}$
for each simulation run.

The individual progenitor galaxies are sampled using $N_\star = 4.15 \times 
10^6$ stellar particles and $N_\mathrm{DM} = 1.0 \times 10^7$ dark matter 
particles for each galaxy, yielding particle masses of $m_\star = 1.0 \times 
10^5 
M_\odot$ and $m_\mathrm{DM} = 7.5 \times 10^6 M_\odot$, respectively. 
Consequently, the ratio between the SMBH mass and the stellar particle mass in 
the simulation sample is between $8500 \leq M_\bullet / m_\star  \leq 85000$, 
which is 
sufficient to realistically study the evolution of SMBH binaries in the 
field of lighter particles \citep{Mikkola1992}. With a sufficiently high 
SMBH-to-stellar particle mass 
ratio the stochastic effects for the SMBH binary evolution are minimized.

\section{Core scouring}\label{section: corescouring}

\subsection{Cusp destruction}\label{section: cusp-destruction}

When the central regions of the merging galaxies coalesce, the 
SMBHs are rapidly driven by the force of dynamical friction to
within the influence radius $r_\mathrm{h}$ of each other. 
We define $r_\mathrm{h}$ as
\begin{equation}\label{eq: rinfl1}
 r_\mathrm{h} = \frac{G M_\bullet}{\sigma_\star^2},
\end{equation}
where $\sigma_\star$ is the line-of-sight stellar velocity dispersion of the 
host galaxy, measured inside the effective radius $R_\mathrm{e}$. Almost 
simultaneously when arriving within $r_\mathrm{h}$, the SMBHs also form a 
gravitationally bound binary. In this section we focus on the simulation sample 
with $\gamma=3/2$ and for these simulations the binaries become bound between 
$t=248$ Myr and $t=275$ 
Myr, with the earlier times referring to the initially more massive SMBHs. 

The subsequent evolution of the binary towards smaller semi-major axis ($a$) 
values is very swift. \cite{Merritt2013} provides an analytical estimate for 
the 
timescale on which the binary transfers energy to the 
surrounding stellar population and a similar computation of the energy transfer 
rate can also be found in \cite{Milosavljevic2001}. The orbital energy of the 
binary, 
defined as
\begin{equation}
E = -\frac{G M_1 M_2}{2 a},
\end{equation}
is transferred into the stellar component on the timescale
\begin{equation}
T_\mathrm{E} = \abs{  \frac{1}{E}  \derfrac{E}{t} }^{-1} \approx 
\frac{\sigma_\star^3}{C G^2 
M_2 \rho_\star}.
\end{equation}
Here $\sigma_\star$ is now the 3D stellar velocity dispersion, $\rho_\star$ is 
the stellar density and $M_1$, $M_2$ are the masses of the SMBHs with $M_2$ 
being 
the less massive SMBH. The 
constant $C$ depends on the primary 
energy transfer mechanism, which is either dynamical friction or three-body 
slingshots. However, 
the value of $C$ is roughly of the order of $C \sim 10$ for both physical 
processes. Note that when 
$r_\mathrm{h} > a$, but the binary is still very wide, the energy transfer 
mechanism is a mixture of the two processes. Computed within the central 
$r<0.19$ kpc of the SMBHs (see \citetalias{Thomas2016}), the energy transfer 
timescales are 
$T_\mathrm{E} \sim 4.3 - 6.9$ Myr for the binaries when $a=r_\mathrm{h}$. This 
timescale is 
small compared to the crossing time of the merger remnant, which is 
$t_\mathrm{cross} = 
R_\mathrm{e}/ \sigma_\star \sim 35$ Myr.

The shrinking SMBH binaries rapidly become hard. We adopt the definition of 
\cite{Merritt2013} for the hardness of the binary:
\begin{equation}
a_\mathrm{h} = \frac{G \mu}{4 \sigma_\star^2} = \frac{q}{1+q}\frac{ 
r_\mathrm{h}}{4},
\label{eq: ah1}
\end{equation}
in which $\mu = M_1 M_2 / (M_1+M_2)$ is the reduced mass and $q=M_2/M_1 < 1$ is 
the mass ratio of the binary. For an equal-mass 
SMBH binary ($M_1 = M_2$, i.e. $q=1$) the hard semi-major axis becomes
\begin{equation}\label{eq: ah}
a_\mathrm{h} = \frac{ r_\mathrm{h}}{8}.
\end{equation}

\begin{figure}
\includegraphics[width=90mm]{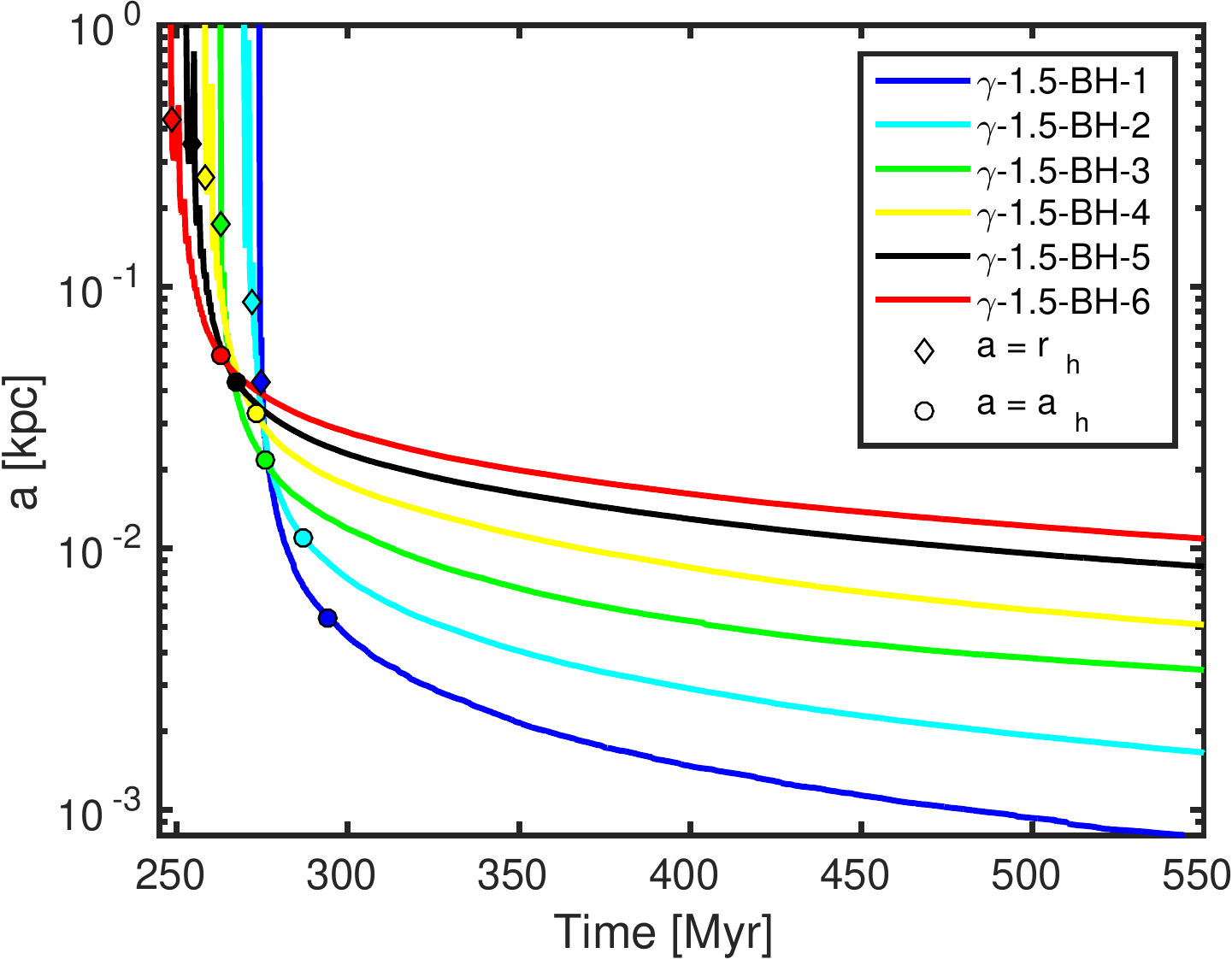}
\caption{The evolution of the semi-major axes of the six SMBH binaries in the 
simulations $\gamma$-1.5-BH-1 to $\gamma$-1.5-BH-6. The SMBH binaries harden 
from 
the influence radius $r_\mathrm{h}$ (open diamonds) to the hard separation 
$a_\mathrm{h}$ (open circles) on a timescale of $\sim 13 - 20$ Myr, which is 
between the 
binary energy transfer timescale $T_\mathrm{E} \sim 4.3 - 6.9$ Myr and the 
crossing time of the host galaxy, $t_\mathrm{cross} \sim 35$ Myr.}
\label{fig: semimajor-axis}
\end{figure}

For our simulation sample, the influence radii of the SMBHs are between $0.09$ 
kpc $ < r_\mathrm{h} < 0.87$ kpc and the hard semi-major axes are in the range 
$0.01$ 
kpc $ < a_\mathrm{h} < 0.11$ kpc. Note that these separations are quite large 
when compared to values commonly found in the literature, the reason being that 
most of the SMBH 
masses in our study are above the values expected from the scaling 
relations between $M_\bullet$ and the host galaxy properties. The SMBH binaries 
shrink from $a=r_\mathrm{h}$ to $a=a_\mathrm{h}$ in $13 - 20$ Myr, which is 
within a factor of $2 - 3$ from 
both the binary energy transfer timescale $T_\mathrm{E} \sim 4.3 -6.9$ Myr and 
the crossing time of 
the host galaxy, $t_\mathrm{cross} \sim 35$ Myr. However, using another popular 
definition 
\begin{equation}
 a_\mathrm{h} = \frac{1}{(1+q)^2} \frac{r_\mathrm{h}}{4}
\end{equation}
for the hard separation (e.g. \citealt{Merritt2006}), the definition differing 
from Eq. \eqref{eq: ah1} by a 
factor of $2$ for an equal-mass binary, we obtain shrinking timescales of $44 - 
55$ Myr. These timescales are somewhat longer than the crossing time of the 
host galaxy. For further discussion of the numerical factors in the definition 
of $a_\mathrm{h}$, see \cite{Merritt2013}. In Fig. \ref{fig: semimajor-axis} we 
plot the evolution of the binary semi-major axes around $a=a_\mathrm{h}$ for the
$\gamma=3/2$ simulation sample.

\begin{figure*}
\begin{center}
	\includegraphics[width=85mm]{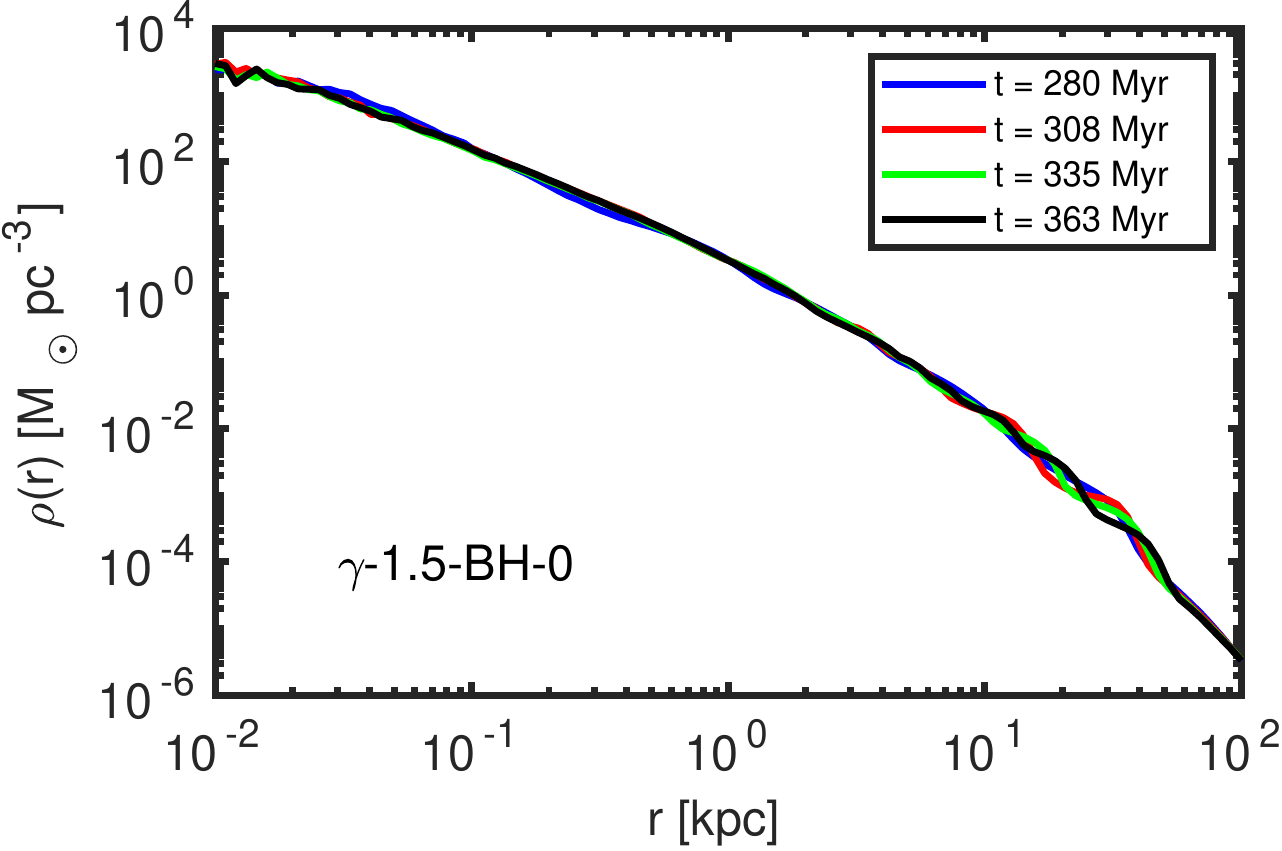}
	\includegraphics[width=175mm]{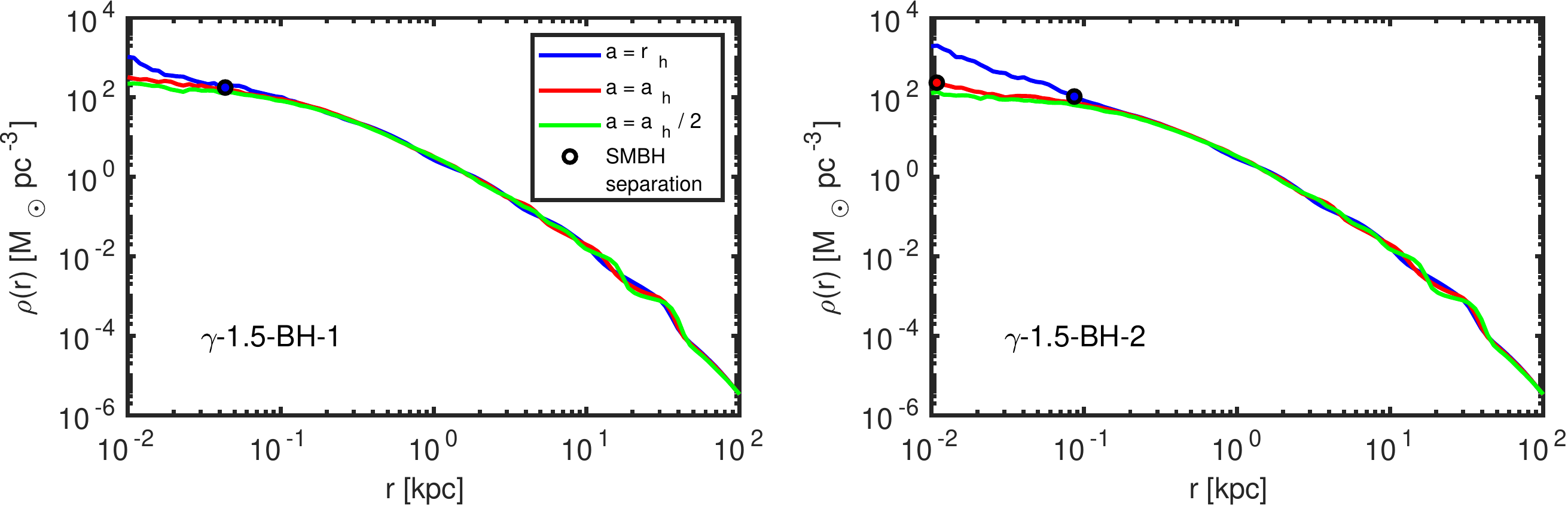}
	\includegraphics[width=175mm]{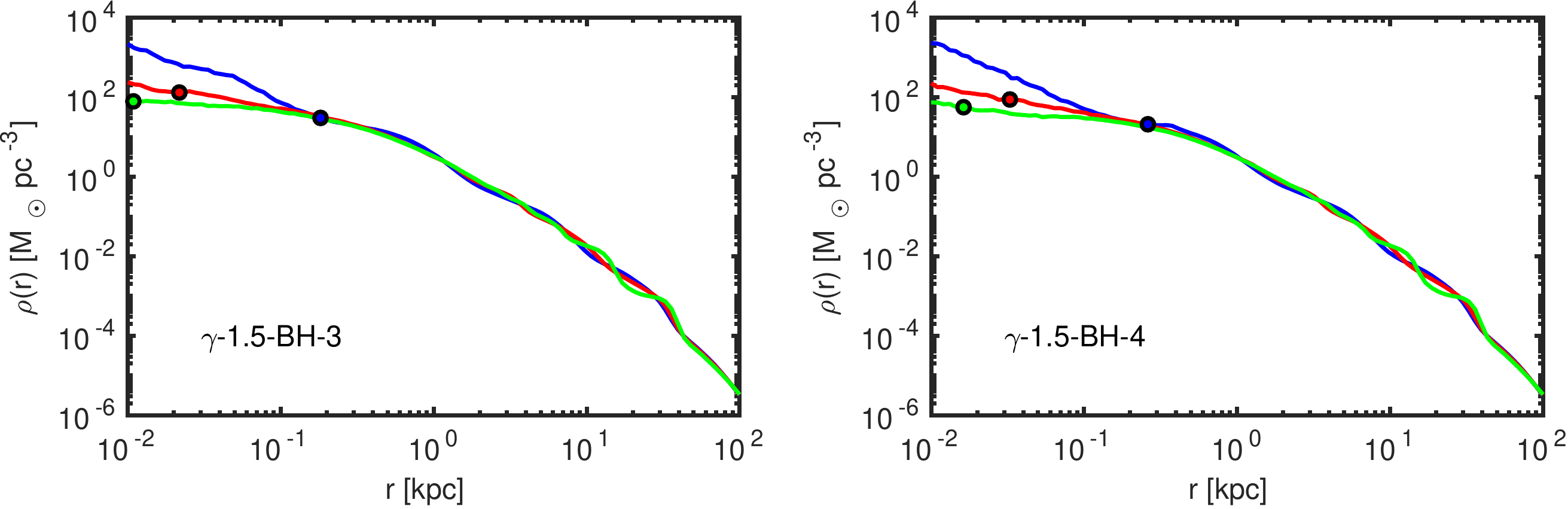}
	\includegraphics[width=175mm]{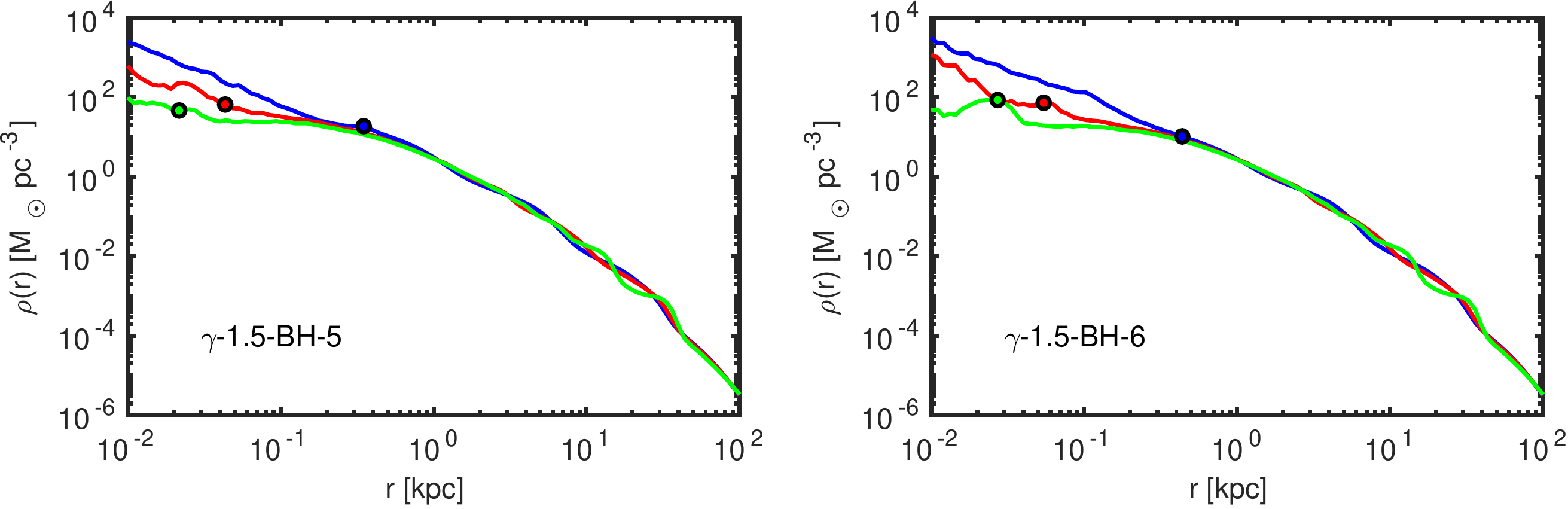}
	\caption{The evolution of the stellar density profiles of the merger 
remnants during the formation and 
hardening phases of the SMBH binaries. The profiles are computed with respect 
to 
the position of one of the SMBHs, 
with this choice remaining fixed throughout the simulations. The SMBH 
separation 
is indicated with an open 
circle if above the plotting range of $10$ pc. The mass deficit and the extent 
of 
the flat central section increases 
systematically with increasing initial SMBH mass.}
\label{fig: density-profiles}
\end{center}
\end{figure*}

The rapid shrinking of the SMBH binaries is accompanied by a sudden decline in 
the stellar density around the binary, transforming the central stellar cusp of 
the host galaxy into a flat low-density core (e.g. 
\citealt{Milosavljevic2001}). 
The evolution of the stellar density profiles in simulations $\gamma$-1.5-BH-0 
to $\gamma$-1.5-BH-6 during 
the core scouring process is presented in Fig. \ref{fig: density-profiles}. At 
the time when the 
SMBHs enter the influence radius $r_\mathrm{h}$ of each other, the central 
stellar cusps of the progenitor galaxies have not yet merged.
In order to study the disruption of the stellar cusps, we compute the radial 
stellar density profiles centered at only one of the SMBHs, 
with this choice remaining fixed throughout the simulations. Calculating the 
profile with respect to the center-of-mass of the SMBHs 
would yield unphysical results for the density of the central region when the 
stellar cusps are still separated by several kpc and 
the center-of-mass of the SMBHs lies between the high-density cusps. In the 
simulation without SMBHs, the center point for the 
density profile computation is found using the shrinking sphere algorithm 
taking 
into account only stars from one of the progenitor galaxies.

The single panel at the top of Fig. \ref{fig: density-profiles} shows that in 
the simulations without SMBHs the stellar density profiles remain
practically unchanged after the coalescence of the stellar cusps of the 
progenitor galaxies. This is consistent with earlier simulation results that
showed that the flattening of the density profiles is expected to be mild in 
collisionless mergers of cuspy progenitors (e.g. \citealt{BK2004}).
The other panels in Fig. \ref{fig: density-profiles} show the evolution of the 
density profiles in the runs with SMBHs, where the separation of the SMBHs
is marked with an open circle for the cases where the separation is larger than 
the plotting range of $10 \ \rm pc$. 

The shrinking binaries displace stellar mass from the central region, with the 
final mass deficit scaling with the mass of the 
SMBH binary \citep{Merritt2006}. This is clearly seen especially in the bottom 
panels with the more massive progenitor SMBHs, for which 
a flattened central section appears in the density profile when the semi-major 
axis of the binary shrinks from 
$a = r_\mathrm{h}$ to $a = a_\mathrm{h}$. The extent of the flat core and the 
amount of displaced 
stellar mass increases systematically as a function of increasing initial SMBH 
mass. Most of the stellar mass to be displaced from the central region (area 
between the blue and red lines in Fig. \ref{fig: density-profiles}) is 
ejected before $a=a_\mathrm{h}$.
The density profiles of the run $\gamma$-1.5-BH-0 without SMBHs are presented 
between $t=280$ Myr and $t=363$ Myr, which coincides 
approximately to the moments of the SMBH binary becoming bound and subsequently 
hard in the simulations containing SMBHs.

\begin{figure}
\includegraphics[width=90mm]{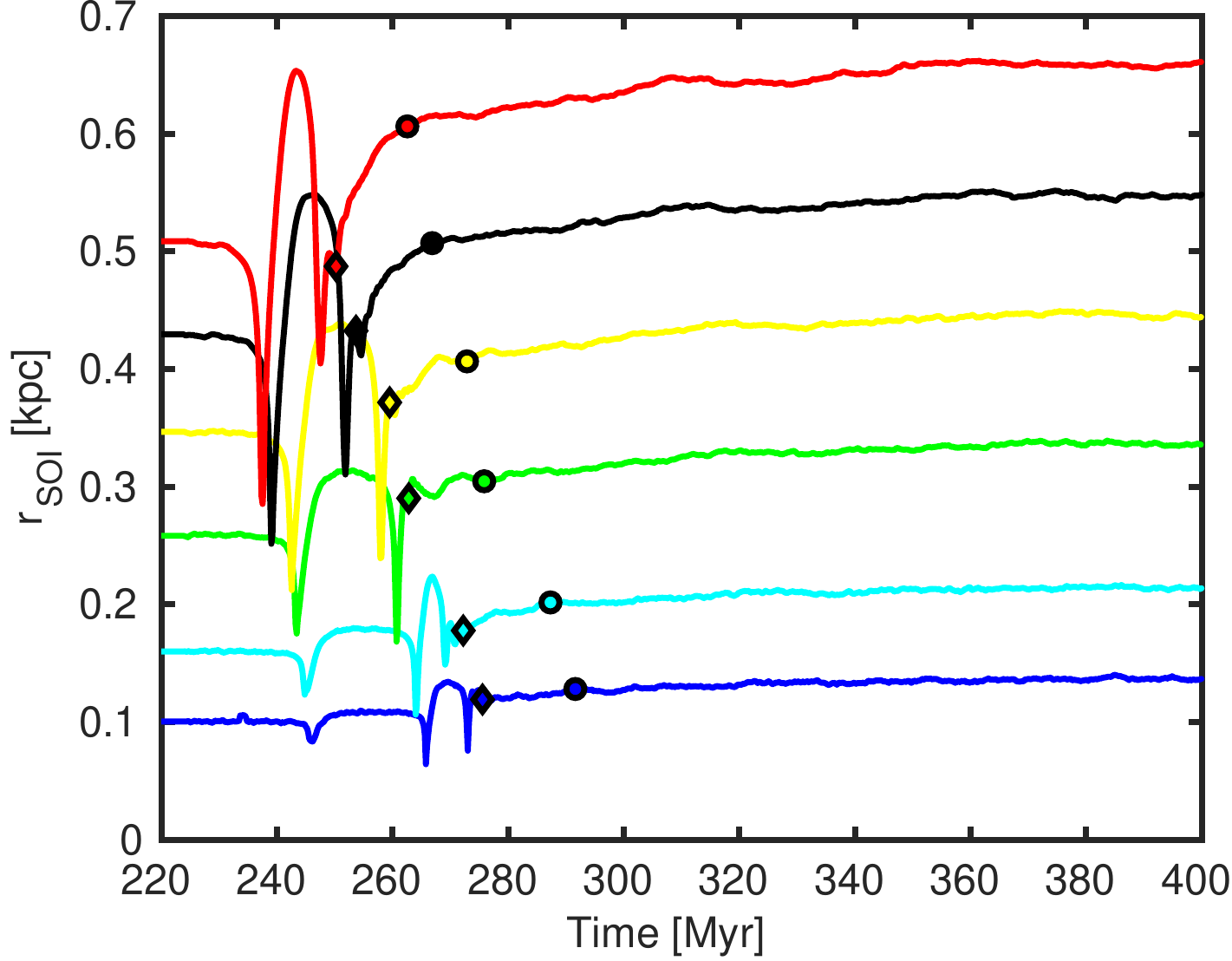}
\caption{The evolution of the sphere-of-influence of an individual SMBH 
($M_\star(r_\mathrm{SOI}) = M_\bullet$) during the formation and hardening 
processes of the SMBH binary. The rapid increase between $a=r_\mathrm{h}$ and 
$a=a_\mathrm{h}$ is due to the strong decline in the stellar density as 
stars are displaced from the central regions by the shrinking binary. This is 
the time when the stellar core forms. 
After the binary becomes hard, the sphere-of-influence $r_\mathrm{SOI}$ only 
increases mildly.}
\label{fig: rsoi}
\end{figure}

The decline of the central stellar density can also be characterized by the 
increasing size of the sphere of influence $r_\mathrm{SOI}$ of the SMBHs.
Here we define $r_\mathrm{SOI}$ as the radius enclosing an stellar mass 
equalling the SMBH mass, $M_\bullet$. The evolution of the sphere of influence 
of an individual SMBH during the core scouring process is presented in Fig. 
\ref{fig: rsoi}. The color coding and symbols describing the different
simulations are the same as in Fig. \ref{fig: semimajor-axis}.  Before the 
SMBHs 
form a gravitationally bound binary, the extent of the sphere of 
influence fluctuates as a response to the changing central stellar density 
caused by the close 
passages of the nuclear stellar cusps. After $a=r_\mathrm{h}$ (open diamond 
symbols), the sphere of influence $r_\mathrm{SOI}$ rapidly increases 
as the stellar density around the shrinking SMBH binary 
strongly declines. This is especially clearly seen in the simulations with more 
massive SMBHs, which result in larger mass deficits. The phase of rapid
evolution of the $r_\mathrm{SOI}$ lasts only until the binary becomes hard 
(open 
circles), after which the rate of evolution of $r_\mathrm{SOI}$ is much 
slower.

\begin{figure*}
	\includegraphics[width=\textwidth]{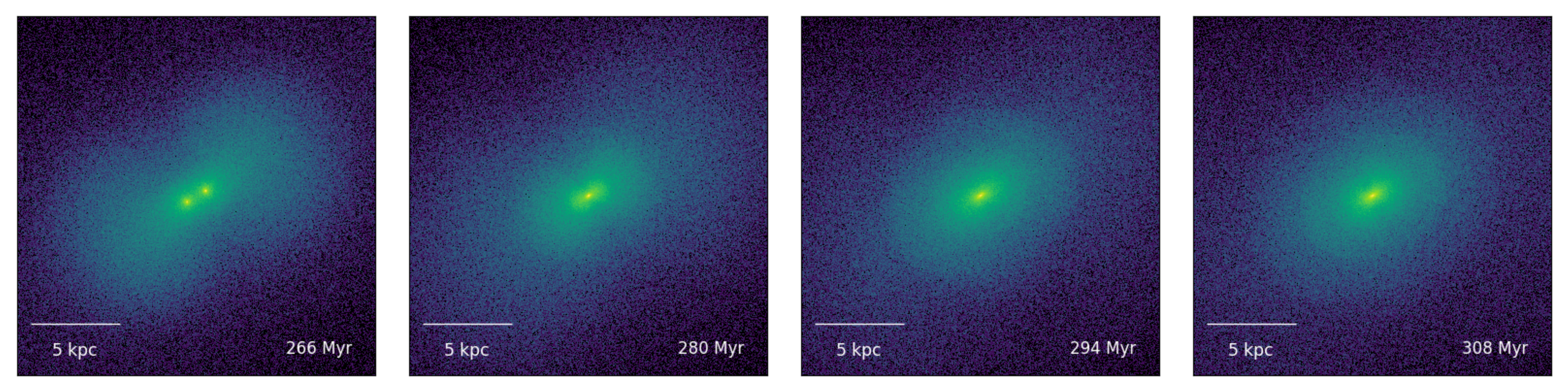}
	\includegraphics[width=\textwidth]{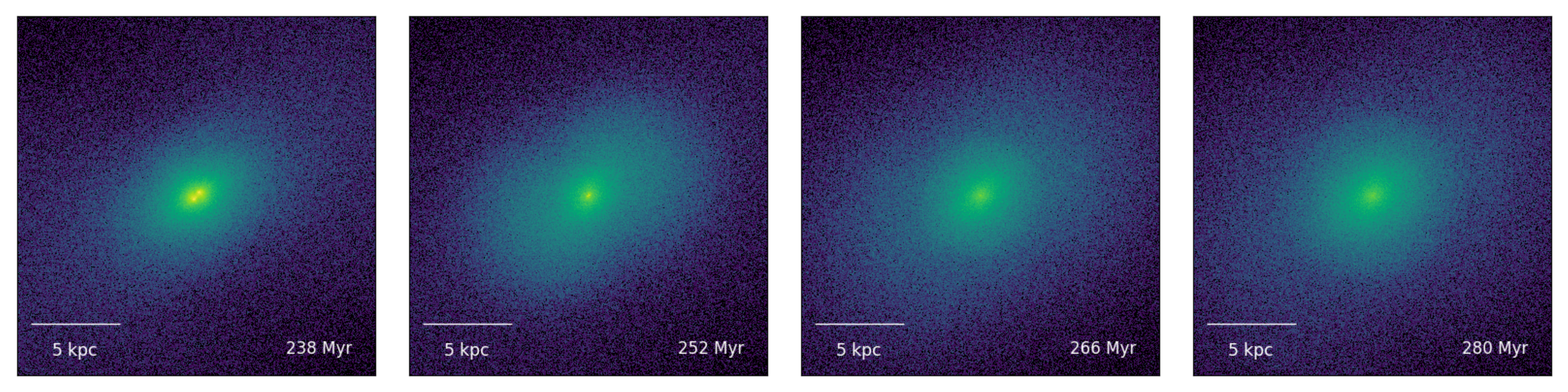}
	\includegraphics[width=\textwidth]{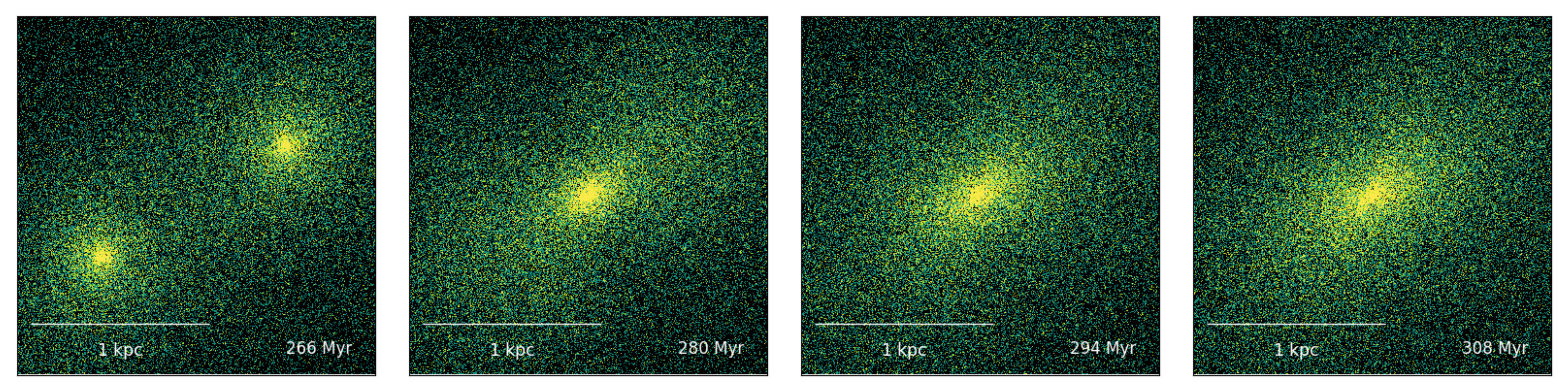}
	\includegraphics[width=\textwidth]{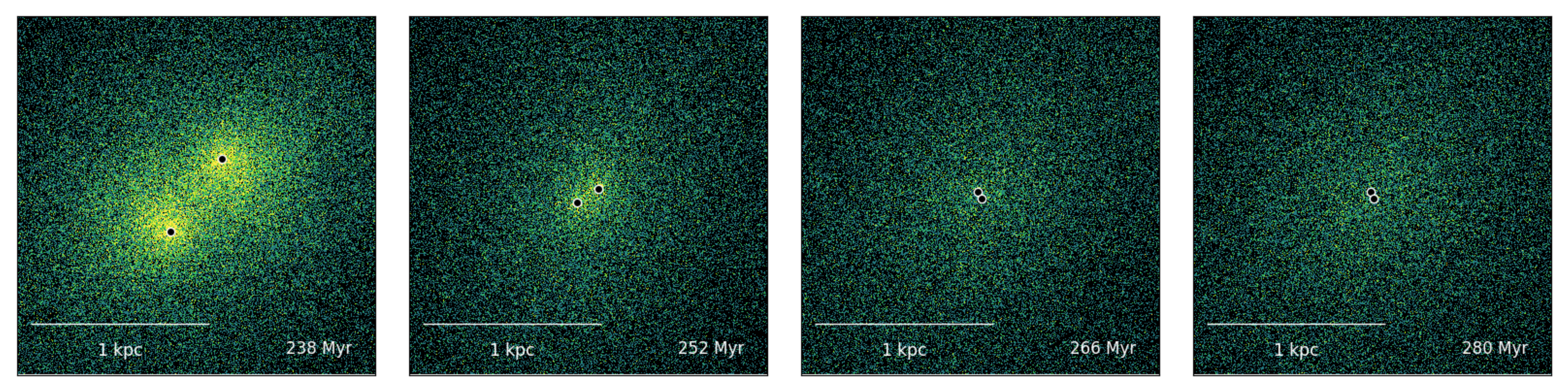}
	\caption{Stellar surface density in the merger simulations at the phase 
of the coalescence of the central density cusps in 
$20$ kpc (top two rows) and $2$ kpc frames (bottom two rows) covering 42 Myr of 
evolution from left to right. The first and third row show 
the simulation $\gamma$-3/2-BH-0 without SMBHs, while rows two and four present 
the simulation $\gamma$-3/2-BH-6 with the most massive SMBHs. 
In the run without SMBHs, the 
central surface density remains high after the formation of a single nucleus, 
although some stars are ejected in outward-moving shells. 
In the run $\gamma$-1.5-BH-6, the central stellar cusps merge somewhat earlier 
due to the additional dynamical friction caused by the 
massive SMBHs. The positions of the SMBHs are indicated by circular markers in 
the panels of the bottom row. During the cusp coalescence, 
the central surface density is significantly reduced on an timescale of the 
order of $10$ Myr seen in the bottom panel moving from 
$t=252 \ \rm Myr$ to $t=266 \ \rm Myr$. A significant fraction of the stars 
are ejected from the central region creating an central low-density core (see 
Fig. \ref{fig: density-profiles}).}
	\label{fig: surfacedensity}
\end{figure*}

The evolution of the two-dimensional surface densities of the merger remnants 
are presented in Fig. \ref{fig: surfacedensity} around the 
time of coalescence of the central stellar cusps. The four rows of Fig. 
\ref{fig: surfacedensity} show the stellar surface density in the 
simulation run $\gamma$-1.5-BH-0 without SMBHs and in the run $\gamma$-1.5-BH-6 
with the most massive SMBHs. 
The spatial extent of the panels is $20$ kpc in the two top rows and $2$ kpc in 
the two bottom rows.
In the run without SMBHs, the central stellar surface density remains high 
after 
cusp coalescence, as seen in the top and third panels of
Fig. \ref{fig: surfacedensity}, with some stars ejected in shells during the 
pericenter passages of the cusps. In the simulations that 
include SMBHs the evolution is qualitatively different, with the central 
stellar 
surface density strongly declining due to the shrinking 
binary (fourth panel in Fig. \ref{fig: surfacedensity}). The stars ejected from 
the central region end up in the outer parts of the merger 
remnant and are visible as a diffuse stellar component in the outer parts of 
the 
merger remnant as can be seen in the rightmost panel of the second row
in Fig. \ref{fig: surfacedensity}.

\subsection{Destruction of radial orbits}\label{section: anisotropy-section}

Next, we study the evolution of the stellar velocity dispersion and the 
velocity 
anisotropy in the center of the merger remnant. Following 
\citetalias{Thomas2016}
we define here the central region as $r<0.19$ kpc.
In spherical coordinates, the components of the velocity dispersion can be 
expressed as
$\sigma_{\mathrm{r}}$, $\sigma_{\mathrm{\theta}}$ and $\sigma_{\mathrm{\phi}}$. 
The two angular 
dispersions can be combined into the tangential velocity dispersion component 
$\sigma_{\mathrm{t}}$ defined as
\begin{equation}
\sigma_{\mathrm{t}}  = \sqrt{\frac{\sigma_{\mathrm{\theta}}^2 + 
\sigma_{\mathrm{\phi}}^2}{2}}.
\end{equation}
The most commonly used parameter to describe the velocity anisotropy structure 
of a stellar system is the velocity anisotropy $\beta$ \citep{Binney2008}, 
defined as
\begin{equation}
\beta = 1 - \frac{\sigma_\mathrm{\theta}^2 + \sigma_\mathrm{\phi}^2}{2 
\sigma_\mathrm{r}^2} = 1 - \frac{\sigma_\mathrm{t}^2}{\sigma_\mathrm{r}^2}.
\end{equation}

The velocity anisotropy $\beta$ is connected to the orbits of the stellar 
population. If all the stellar orbits are radial, $\sigma_\mathrm{t} = 0$ and 
$\beta=1$. On the other hand, for a stellar population on purely circular 
orbits 
$\sigma_\mathrm{r} = 0$ and $\beta = -\infty$. SMBH binaries can interact 
strongly with stars that have small pericenter distances, corresponding to low 
angular momenta. 
Consequently, SMBH binaries are expected to primarily eject stars on radial 
orbits and produce a tangentially biased ($\beta<0$) stellar population in the 
core region. The main aspects of this model have been verified both by 
numerical 
simulations \citep{Quinlan1997, Milosavljevic2001} and observations (e.g. 
\citealt{Thomas2014}).

The evolution of the central radial and tangential velocity dispersions 
($\sigma_\mathrm{r}$, $\sigma_\mathrm{t}$) as well as the velocity anisotropy 
$\beta$ is presented for the $\gamma=3/2$ simulation sample in Fig. \ref{fig: 
dispersions}. Initially, the radial and tangential 
velocity dispersions are equal in the vicinity of the SMBHs resulting in $\beta 
\sim 0$ before the SMBHs form a bound binary. As the binary forms and shrinks 
from the influence radius $a=r_\mathrm{h}$ towards the hard separation $a = 
a_\mathrm{h}$, both the 
central radial and tangential velocity dispersions increase as more mass is 
brought into the center of the galaxy. The velocity anisotropy remains still 
$\beta \sim 0$ in this evolutionary phase.

\begin{figure}
\includegraphics[width=90mm]{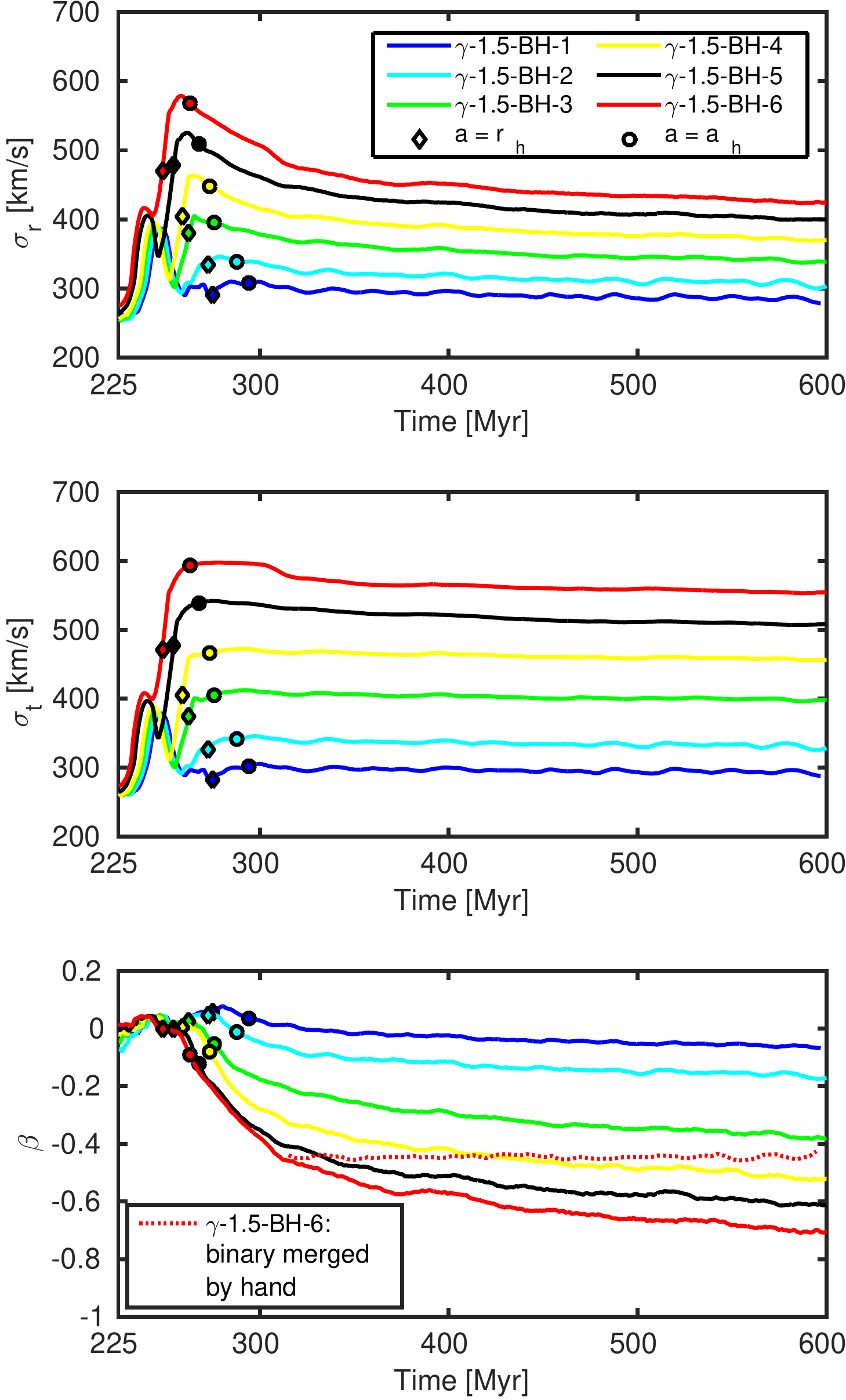}
\caption{The evolution of the central ($r<0.19$ kpc) radial and tangential 
components of the stellar velocity dispersion 
$\sigma_\mathrm{r},\sigma_\mathrm{t}$ as well as the central 
velocity anisotropy $\beta$. Both $\sigma_\mathrm{r}$ and $\sigma_\mathrm{t}$ 
increase until $a=a_\mathrm{h}$. The hard binary ejects stars on radial orbits, 
causing the radial velocity dispersion to decline while the stars on more 
circular orbits remain unaffected with $\sigma_\mathrm{t} \sim$ constant. The 
slingshot process results in a tangentially biased velocity 
structure, with $\beta<0$ in the center of the merger remnants.}
\label{fig: dispersions}
\end{figure}

The radial velocity dispersion $\sigma_\mathrm{r}$ reaches its maximum value 
almost coincidentally with the SMBH binary becoming hard. After this, the 
tangential velocity dispersion remains roughly constant while the radial 
velocity dispersion begins to decline. The decline is stronger for galaxies 
with 
higher 
initial SMBH masses. Consequently, a tangentially biased region with $\beta < 
0$ 
is formed 
in the 
central region of the galaxy. We performed an additional test simulation to 
confirm that the SMBH binary scouring indeed causes the decline of the central 
$\beta$, and that it is not due to numerical effects. The SMBHs of 
simulation 
$\gamma$-1.5-BH-6 were merged by hand when 
$\beta = -0.4$ and the run was restarted. The evolution of the velocity 
anisotropy in this run is 
presented with a dotted line in the bottom panel of Fig. \ref{fig: 
dispersions}. The decline of $\beta$ ceases immediately if the SMBHs are merged 
by hand, 
confirming that the hard, shrinking SMBH binaries cause the decline of the 
central velocity 
anisotropy.

A star which experiences a strong interaction with the SMBH binary is 
typically ejected with a velocity $v_\star$ comparable to the circular orbital 
velocity 
of the binary:
\begin{equation}
 v_\star \sim V_\mathrm{bin} = \left( \frac{2 G M_\bullet}{a} \right)^{1/2},
\label{eq:v_binary}
\end{equation}
even though the distribution $f(v_\star)$ of the ejection velocities is very 
broad 
(\citealt{Valtonen2006, Merritt2013}). The three-body slingshots become 
important when the typical ejected star is able to completely escape the host 
galaxy.
A commonly used criterion for escape from a stellar system in N-body studies is 
$v_\mathrm{esc} = 2 \sqrt{3} \sigma_\star$, simplified from $\langle 
v_\mathrm{esc}^2 \rangle = 12 \langle \sigma_\star^2\rangle$, which assumes 
only 
the virial theorem and an isotropic velocity dispersion tensor 
\citep{Spitzer1987}. For the merger 
remnants in our simulations, this criterion yields $v_\mathrm{esc} \sim 1000$ 
km/s. 
In order to study when this occurs in our simulations, we insert the 
definitions 
of the influence radius and the hard separation from Eqs. \eqref{eq: rinfl1} 
and 
\eqref{eq: ah} into the definition of the circular binary velocity Eq. 
\eqref{eq:v_binary}. 
We obtain the typical ejection velocities 
\begin{equation}
 v_\star  = 
 \begin{cases}
\sqrt{2} \sigma_\star \sim 410 \hspace{0.1cm} \mathrm{ km/s}, & a=r_\mathrm{h}\\
4 \sigma_\star \sim 1160 \hspace{0.1cm} \mathrm{ km/s}, & a=a_\mathrm{h}.\\
\end{cases}
\end{equation}

At $a = r_\mathrm{h}$, the mean stellar ejection velocity is still low and the 
slingshot mechanism is ineffective.
The hard binaries eject stars at typical velocities of $v_\star = 4 
\sigma_\star 
> 
v_\mathrm{esc}$, thus the slingshot mechanism is very effective. When the 
typical 
ejection velocity equals the escape velocity, $v_\star = v_\mathrm{esc}$, the 
semi-major axis of the binary is $a = r_\mathrm{h}/6 \lesssim a_\mathrm{h}$. This 
is 
consistent with the radial velocity dispersion $\sigma_\mathrm{r}$ reaching its 
maximum value a few Myr before the binaries become hard, as can be seen in the 
top panel of 
Fig. \ref{fig: dispersions}. After this point, the binary efficiently destroys 
radial orbits. As stars on more circular orbits cannot interact strongly 
with the SMBH binary (pericenter distance $p \gg a$), the tangential velocity 
dispersions of the merger remnants remain relatively unaffected.

The results in Sections \S \ref{section: cusp-destruction} and \S \ref{section: 
anisotropy-section} indicate that the formation of 
the low-density core region and the development of the tangentially biased 
velocity dispersion occur mainly at different times during
the evolution of the merger remnant. Most of the stellar mass to be displaced 
from the core region is removed before the SMBH binary
becomes hard at $a=a_\mathrm{h}$, whereas most of the central negative velocity 
anisotropy $\beta$ is built up after this stage. 
In addition, the timescales of the two processes are very different. The 
central 
stellar density declines on a timescale of the order of
tens of Myr, which is below the crossing time of the merger remnant. On the 
contrary, the build-up of the tangentially biased velocity 
distribution occurs on a timescale of $\sim$ a few hundred Myr, which is 
several 
times the crossing time of the merger remnant.

\subsection{Binary evolution after core formation}

Using \ketju, we include in this study PN corrections up to order 2.5 in the 
equations of 
motion of the SMBHs. The PN corrections, and thus the gravitational 
wave (GW) emission effects are still negligibly small when $a \sim 
a_\mathrm{h}$, excluding the possible but very rare head-on collisions of 
SMBHs. 
Eventually, depending on the SMBH binary mass and eccentricity, the PN 
radiative 
loss terms in the equations of motion become important. This occurs at the 
latest 
when $a \sim a_{\mathrm{GW}} \sim 0.01 \times a_\mathrm{h}$ 
\citep{Quinlan1996}, 
yielding $a_{\mathrm{GW}} \sim 0.1 - 1.1$ pc for the binaries in our sample. 
The 
binary emits the rest of the 
orbital angular momentum and energy away as gravitational waves, resulting in a 
SMBH merger and the 
formation of a single SMBH.

Averaged over the orbital period, the rates of change of the 
binary semi-major axis and eccentricity are \citep{Peters1963}
\begin{flalign}
&\left\langle \derfrac{a}{t} \right\rangle_\mathrm{GW} = - \frac{64}{5} 
\frac{G^3 M_{1} 
M_{2}    (M_{1}+M_{2})}{c^5 a^3} \frac{1+\frac{73}{24} e^2 + \frac{37}{96} 
e^4}{(1-e^2)^{7/2}} \nonumber \\
&\left\langle \derfrac{e}{t} \right\rangle_\mathrm{GW} = - \frac{304}{15} e 
\frac{G^3 
M_{1} M_{2}
    (M_{1}+M_{2})}{c^5
a^4} \frac{1+\frac{121}{304} e^2 }{(1-e^2)^{5/2}} \label{eq: peters}
\end{flalign}
assuming that the binary evolution is driven purely by GW emission in the 
Post-Newtonian order PN2.5. Here $M_1,M_2$ are again the masses of the SMBHs, 
$a$ is the semi-major axis and $e$ is the eccentricity of the binary. An 
important property of Peters' formulas \eqref{eq: peters} is the extreme 
dependence of the GW-driven binary evolution on the 
binary eccentricity. Both $\langle \dot{a}\rangle$ and $\langle \dot{e}\rangle$ 
diverge in the limit $e \to 1$.

Most of the simulations are run until the SMBHs merge. However, as our main 
priority 
lies in studying the core scouring process in the simulated galaxies we do not 
continue simulation runs with low-eccentricity binaries beyond $t=2$ Gyr.  
The maximum possible SMBH coalescence timescale in our simulation sample is 
$t_{\mathrm{coal}} \sim 6$ Gyr, assuming the lowest hardening rate among the 
simulations (run $\gamma$-1.0-BH-6) and a circular binary with $e=0$ 
\citepalias{Rantala2017}.

\subsection{Final surface brightness profiles}\label{section: surfacebrightness}

\begin{figure*}
\includegraphics[width=\linewidth]{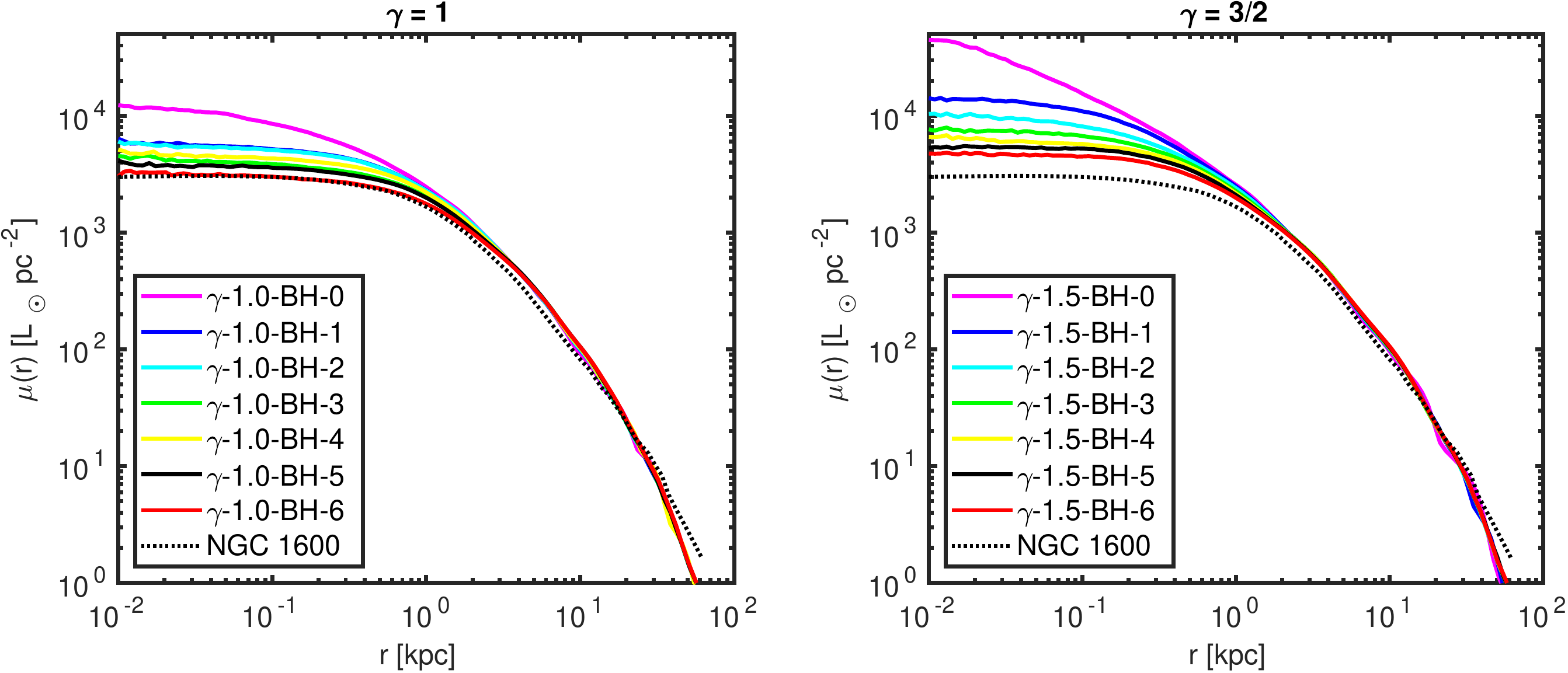}
\caption{The final surface brightness profiles (solid lines) of the merger 
simulations with $\gamma=1$ (left panel) and $\gamma=3/2$ (right panel). We 
assume a constant stellar mass-to-light ratio of $M/L_\odot = 4.0$. In both 
cases 
the central surface brightness decreases and the core size increases with 
increasing initial SMBH mass. For comparison, we show the observed profile of 
NGC 1600 
\citepalias{Thomas2016} as the black dotted line. 
We find that the agreement is best for simulations 
for which the final SMBH mass is similar to the SMBH mass inferred for NGC 
1600.}
\label{fig: surfacebrightness}
\end{figure*}

The final surface brightness profiles $\mu(r)$ of the merger remnants are 
presented in Fig. \ref{fig: surfacebrightness}. As in \citetalias{Thomas2016}, 
we assume a constant stellar mass-to-light ratio of $M_\star/L = 4.0$ in order 
to 
compare the simulated surface density profiles with actual observations. The 
profiles are azimuthally 
averaged over $100$ random viewing angles. The outer parts of the simulated 
merger remnants appear very similar, as expected from the Dehnen profile 
initial conditions and identical merger orbits. The surface brightness of the 
simulated galaxies falls below the observed profile of NGC 1600 at $r \sim 30$ 
kpc.

Studying the central parts of the simulated merger remnants, the surface 
brightness profiles in the runs that include SMBHs turn almost flat at $0.1$ 
kpc 
$\lesssim 
r_{\mathrm{b}} \lesssim 1$ kpc. The profiles of the merger remnants without 
SMBHs remain cuspy until the very center. We also immediately see that the 
central 
surface brightness is lower for the $\gamma = 1$ initial conditions compared to 
the $\gamma = 3/2$ runs with the same SMBH mass. In addition, the central 
surface brightness decreases and the core size (extent of the flat section) 
increases as the initial SMBH mass increases in both progenitor galaxy samples. 
This is consistent with the mass deficit arguments in the 
literature, which states that the mass deficit is proportional to the total 
mass of the SMBH binary (e.g. \citealt{Merritt2006}). 

The agreement with the observations is best in the simulations with the final 
SMBH mass similar to $M_\bullet = 1.7 \times 10^{10} M_\odot$ inferred for NGC 
1600. However, there is some ambiguity here because of the assumed stellar 
mass-to-light ratio $M_\star/L  = 4.0$ \citep{vanDokkum2017}. With larger 
values 
for the 
mass-to-light 
ratio $M_\star/L 
>4.0$, simulated galaxies with lower $M_\bullet$ or steeper 
initial stellar density profiles could also match the central surface 
brightness 
of NGC 1600. However, with $M_\star/L >4.0$, the outer parts of the simulated 
galaxies 
would end up too dim compared to the observations. As NGC 1600 is in a group 
environment, the
galaxy has most likely experienced several minor mergers that have puffed up 
and 
brightened the 
outer parts of the galaxy, 
as opposed to our isolated merger models (e.g. \citealt{Hilz2012, Hilz2013}). 
Based solely on surface brightness data, we cannot exactly state which one of 
the simulated galaxies most closely resembles the observed NGC 1600, and 
additional velocity data is required for this.

\subsection{Final velocity anisotropy profiles}

\begin{figure*}
\includegraphics[width=\linewidth]{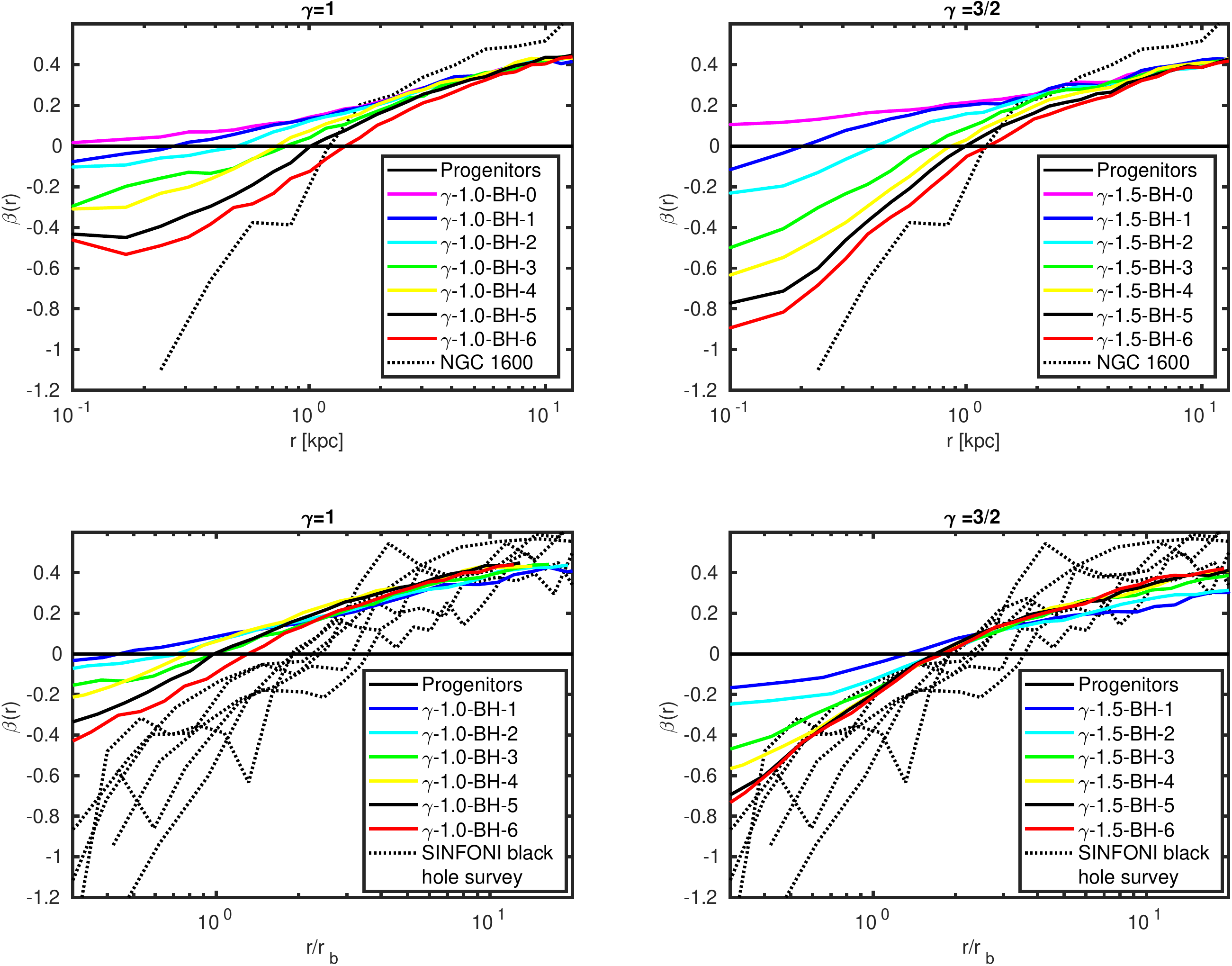}
\caption{The velocity anisotropy profiles $\beta(r)$ (thick solid lines) in the 
merger simulations with $\gamma=1$ (left column) and $\gamma=3/2$ (right 
column). The isotropic progenitor galaxies are presented as thin horizontal 
solid lines with $\beta(r)=0$. 
The dynamical model for NGC 1600 \citepalias{Thomas2016} is shown as the 
dotted line in the two top panels. In the central region ($r<1$ kpc), $\beta$ 
becomes increasingly more negative with increasing initial SMBH mass, 
indicating 
an increasingly more tangentially biased stellar population. The mergers with 
more 
cuspy initial stellar density profiles develop a steeper $\beta$ profile. The 
simulations with the largest initial SMBH masses agree best with the dynamical 
model of NGC 1600. We see a trend in which the steeper initial density cusp 
runs 
produce
better agreement with the inferred dynamical model for NGC 1600.
The two bottom panels present the $\beta$ profiles of 
the merger simulations with the radial coordinate scaled by the break radius 
$r_\mathrm{b}$ of the surface brightness profiles. The models of the 
observed galaxies from the SINFONI black hole survey \citep{Thomas2014, 
Saglia2016} are plotted as dotted lines. 
Again, increasing the 
initial SMBH mass and the steepness of the slope for initial stellar density 
profile produces merger 
remnants in better agreement with the dynamical models of the observed 
galaxies.}
\label{fig: anisotropy}
\end{figure*}

The final velocity anisotropy profiles $\beta(r)$ of the simulated galaxies are 
presented in Fig. \ref{fig: anisotropy}. The velocity anisotropy is computed in 
radial bins, with the bin 
width $\Delta r = 0.19$ kpc motivated by the spatial resolution of the 
spectral data in \citetalias{Thomas2016}. For comparison, the isotropic $\beta$ 
profiles of the progenitor galaxies are also shown as horizontal lines.
The top panels 
compare the $\beta(r)$ of the simulated galaxies to the dynamical model of NGC 
1600 \citepalias{Thomas2016}. Without the inclusion of SMBHs, the velocity 
structure of 
the mergers remnants of both the $\gamma=1$ and $\gamma=3/2$ samples is close 
to 
isotropic in the center of the galaxy. 

In the outer parts, the stellar population is dominated by radial 
orbits ($\beta \sim 0.40$-$0.45$ at $r=10$ kpc) which is typical for 
collisionless mergers \citep{Rottgers2014}. The outer parts of the merger 
remnants with SMBHs are almost identical to the remnants without SMBHs. The 
dynamical model of NGC 1600 is more radially biased at radii beyond $r \gtrsim 
3$ kpc, with 
($\beta \sim 0.5$-$0.6$ at $r=10$ kpc). The smaller amount of stellar material 
on radial 
orbits in the outer parts can be explained by the lack of minor merges in our 
isolated major merger simulation, as already discussed in the previous 
section. However, it should be noted that the velocity structure of the stellar 
population in the outer parts of the simulated merger remnants show still 
qualitatively 
similar trends to the dynamical model of NGC 1600, at least when compared to 
the 
isotropic progenitor galaxies.

The central parts ($r \lesssim 1$ kpc) of the simulated merger remnants have a 
tangentially biased stellar population ($\beta<0$). There is a clear trend of 
increasingly more negative $\beta$ with the increasing initial SMBH mass. In 
addition, more cuspy initial stellar profiles produce steeper $\beta$ profiles 
in the core region, consistent with the results of \cite{Bortolas2018}. The dynamical model of NGC 1600 has an even steeper 
$\beta$ profile near the center when compared to the simulations. 
This suggests that the initial stellar density profiles for the progenitors of 
NGC 1600
were even steeper than $\gamma=3/2$. We note that the final central values of 
$\beta$ depend on 
the chosen isotropic initial conditions. A second generation of mergers would 
already have $\beta<0$ 
initially, leading to an increasingly tangential central stellar population. In 
addition, other 
physical processes such as the adiabatic growth of the single central SMBH may 
decrease $\beta$ as well (e.g. \citealt{Goodman1984, Thomas2014}).

The bottom panels of Fig. \ref{fig: anisotropy} present the same simulated 
anisotropy profiles as in the top panels, but the radial coordinate is now 
scaled with 
the break radius $r_\mathrm{b}$ of the surface brightness profile. The 
determination of the break radius $r_\mathrm{b}$ is discussed 
in \S \ref{section: coresersic}. The $\beta$ profiles of dynamical models 
of six observed core galaxies from the SINFONI black hole survey 
\citep{Thomas2014, 
Saglia2016} are 
presented for comparison. Observed massive elliptical galaxies with cores have 
remarkably similar velocity anisotropy profiles when scaled with the core 
radius $r_\mathrm{b}$. In the outer parts ($r/r_\mathrm{b}>2$-$3$), the 
$\beta$ 
profiles of the simulated merger remnants and the galaxies from the SINFONI 
black hole survey now 
agree very well. At even larger radii $(r/r_\mathrm{b} > 10)$ the 
simulated 
profiles agree with the lower end of the distribution of observed profiles. In 
the core region ($r/r_\mathrm{b} < 1$), the profiles of the simulated merger 
remnants coincide with the upper end of the 
observed anisotropies in the SINFONI black hole survey. Increasing the initial 
SMBH mass and 
selecting a steeper stellar density profile slope results in merger remnants in 
better agreement
with the dynamical models of the galaxies from the SINFONI black hole survey. 
Again, we note that the assumption 
of a single generation of galaxy mergers with isotropic initial conditions 
affects the results discussed here. 
An additional generation of mergers would start with $\beta<0$ in the center 
and 
$\beta>0$ in the outer parts of the galaxies, potentially 
leading to a better agreement with the observed galaxies.

In summary, the results of our surface brightness profile and 
velocity anisotropy analysis support the SMBH mass $M_\bullet = 
1.7\times10^{10} 
M_\odot$ obtained 
by the dynamical modeling of \citetalias{Thomas2016} and suggest steep initial 
stellar profiles with $\gamma \geq 3/2$ for the progenitor galaxies of NGC 1600.

\subsection{2D kinematics maps}

\begin{figure*}
\begin{center}
\includegraphics[width=140mm]{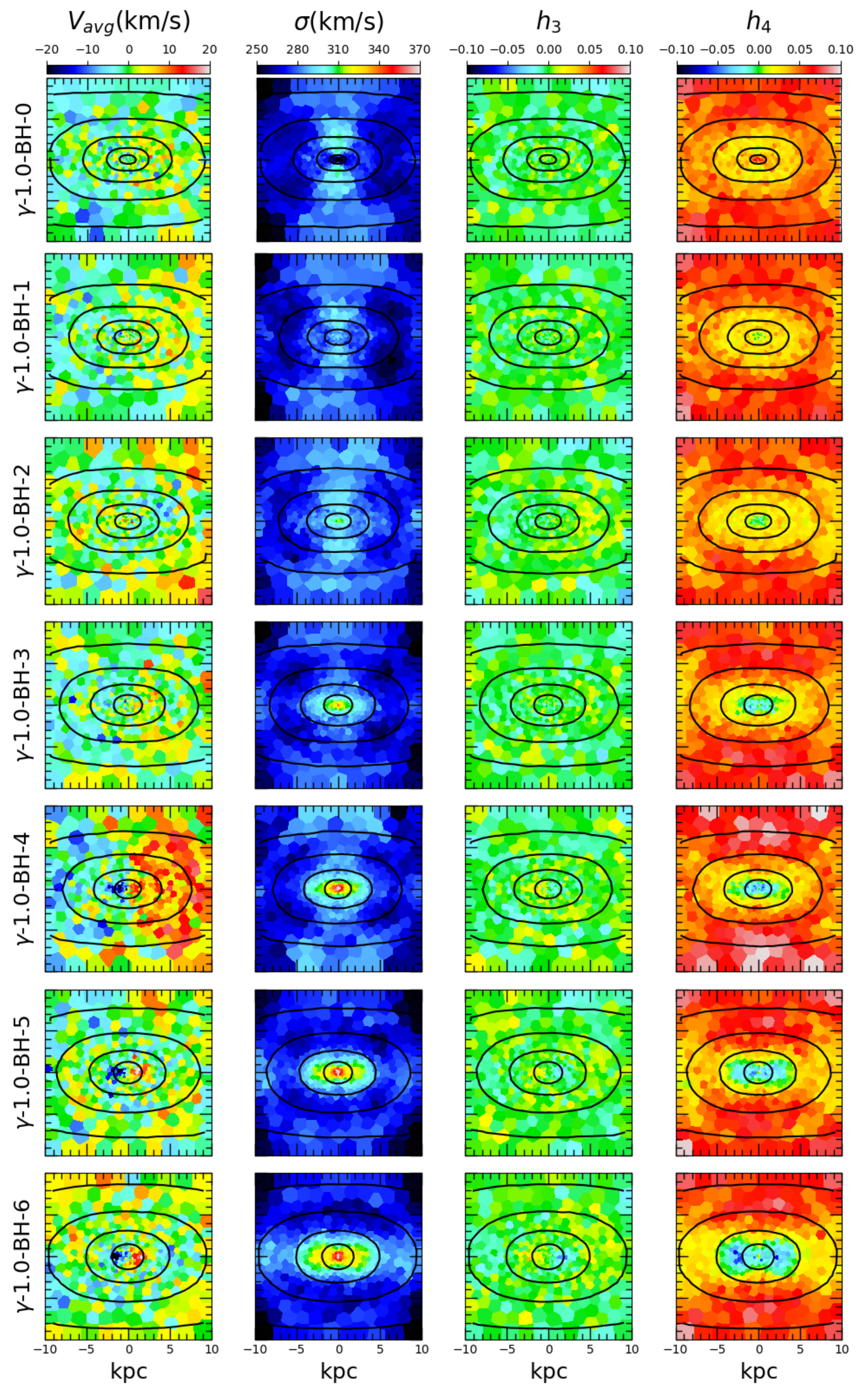}
\caption{Two-dimensional stellar kinematic maps of the simulations with 
$\gamma=1$, the SMBH mass is increasing from top to bottom. Shown are the 
line-of-sight velocity $V_\mathrm{avg}$, the stellar velocity dispersion 
$\sigma$, $h_3$ and $h_4$ 
from left to right. Luminosity isophote 
contours are overlaid with a spacing of one magnitude. The systems with higher 
SMBH masses are developing a 
rotating, 
high-dispersion decoupled core with clearly anti-correlated $h_3$ and 
$V_\mathrm{avg}$, as well as a negative $h_4$.}
\label{fig: ifu_1}
\end{center}
\end{figure*}

\begin{figure*}
\begin{center}
\includegraphics[width=140mm]{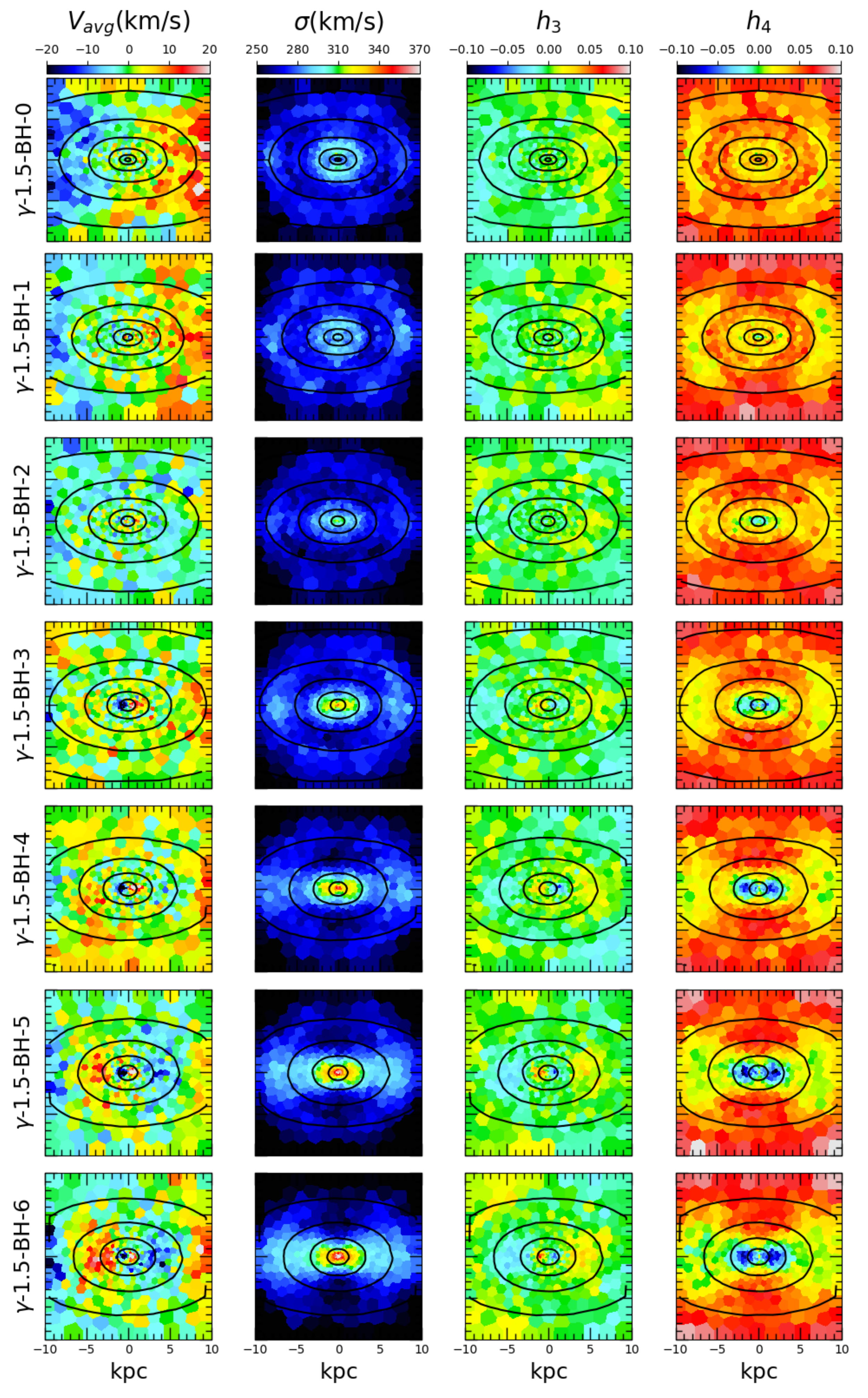}
\caption{The same figure as Fig. \ref{fig: ifu_1} but for simulations with 
$\gamma$=3/2. 
For this simulation sample the rotating central feature becomes also more 
prominent for higher initial SMBH masses. 
There are even indications in the line-of-sight velocity $V_\mathrm{avg}$ and 
$h_3$ maps for more complex 
kinematic subsystems in the merger remnants $\gamma$-1.5-BH-4 to 
$\gamma$-1.5-BH-6.}
\label{fig: ifu_2}
\end{center}
\end{figure*}

We show in Figures  \ref{fig: ifu_1} and \ref{fig: ifu_2} the two dimensional
kinematic maps for the $\gamma = 1$ and $\gamma = 3/2$ simulation samples, 
respectively. 
The maps are constructed similarly to what an observer would see if the merger 
remnants were observed with an Integral Field Unit spectrograph (see 
\citealt{Naab2014}).
Following the same procedures 
used in analyzing observational data, each panel is divided into "spaxels" with 
the same signal-to-noise, using the Voronoi tessellation algorithm 
\citep{Cappellari2003}. In our case the signal-to-noise is represented by the 
number of particles in each bin. For each row in the two figures, the four 
panels show from left to right, the average line-of-sight velocity 
$V_\mathrm{avg}$, the line-of-sight 
velocity dispersion $\sigma$, 
and the $h_3$ and $h_4$ 
parameters, which represent the skewness and kurtosis of the line-of-sight 
(LOS) velocity distribution \citep{vanderMarel1993, Gerhard1993}. These 
parameters are calculated by fitting the 
histogram 
of LOS velocities in every spaxel with the following modified Gaussian function:
\begin{equation}
f ( v )  = I_0 \, e^{ - y^2 / 2 } (1+h_3 \, H_3 (y)+h_4 \, H_4 (y))
\end{equation}
where $y = (v - V_{\rm avg}) / \sigma$ and $I_0$ is a normalization constant. 
Here $H_3$ and $H_4$ are the Hermite polynomials of third and fourth order:
\begin{equation}
H_3(y) =  (2 \, \sqrt{2} \, y^3 - 3 \, \sqrt{2} \, y) / \sqrt{6}
\end{equation}
\begin{equation}
H_4(y) =  (4 \, y^4 - 12 \, y^2 + 3) / \sqrt{24},
\end{equation}

Before performing this analysis, the galaxies are oriented using the reduced 
inertia tensor so that their major axis is on the x axis of the figure, and the 
LOS velocities are centered so that the spaxels within the central kpc in 
projection have zero average velocity. We also plot luminosity contours with a 
spacing of one magnitude on top of each velocity map.

In the merger 
remnants without SMBHs, the 
central regions show a positive $h_4$ peak. This is typical for nearly 
isotropic 
galaxies with SMBHs and not too 
steep central stellar cusps \citep{Baes2005}. However, in simulations including 
SMBHs, we find systematically more negative $h_4$ values as a function of 
increasing SMBH mass (see bottom rows of Figs. \ref{fig: ifu_1} and \ref{fig: 
ifu_2}). The size of the negative $h_4$ region also grows 
with increasing SMBH mass and increasing size of the sphere of influence. The 
lower values of $h_4$ likely result from the missing weakly bound stars on 
low-angular-momentum orbits. 

In the velocity maps a central kpc-scale, decoupled rotating region becomes 
increasingly more visible as the SMBH mass increases. There are even 
indications of more complex kinematic subsystems in the merger remnants 
$\gamma$-1.5-BH-4 to 
$\gamma$-1.5-BH-6. As can be seen in the bottom panels of Fig. \ref{fig: ifu_2} 
a large ($\sim 5$ kpc) region counter-rotating to the central rotating region 
becomes 
apparent in the simulations containing the more massive SMBHs. The $h_3$ map 
shows an anti-correlation to the LOS velocity map, which is commonly observed 
in 
rotating systems (e.g. \citealt{Bender1994, Krajnovic2011, Veale2017}). 
Finally, 
the stellar velocity 
dispersion increases also because of the increased mass in the central region 
for larger SMBH masses. 

In conclusion, Fig. \ref{fig: ifu_1} \& Fig. \ref{fig: ifu_2} clearly 
demonstrate that the effects of black hole scouring are indeed observable in 
the 
kinematic maps of quiescent merger remnants.

\section{Scaling relations}
\label{section:scaling_relations}

\subsection{Observed and simulated galaxy samples}\label{section: 
scaling-samples}

In \citetalias{Thomas2016} tight scaling relations are presented 
between the observed galaxy core sizes (or break radii) $r_\mathrm{b}$, the 
SMBH 
masses $M_\bullet$ and the spheres of influence of the SMBHs $r_\mathrm{SOI}$. 
We 
perform a similar analysis for our simulated merger remnants. In addition, we 
use a simple core scouring model to interpret the results of our merger 
simulations.

It should be noted that while our merger simulation sample consists of 
single-mass 
remnants with $M_\star = M_\star^\mathrm{NGC1600} = 8.3 \times 10^{11} 
M_\odot$, 
the 
scaling relations of 
\citetalias{Thomas2016} are obtained from a broader core galaxy sample. The 
core 
galaxies in the \citetalias{Thomas2016} sample have total stellar masses in the 
range from $M_\star = 5.0\times10^{10} M_\odot$ to $2.0\times10^{12} M_\odot$, 
with a mean stellar 
mass of $\langle M_\star \rangle = 5.8 \times10^{11} M_\odot$ 
\citep{Rusli2013, Rusli2013b, Saglia2016}. The mean effective radius is 
approximately
$\langle R_\mathrm{e} \rangle = 8.5$ kpc. The velocity dispersions of the 
observed sample 
are in the range $168 <\sigma_\star<380$ km/s with a mean of $\langle 
\sigma_\star \rangle = 289$ km/s.
This agrees reasonably well with the velocity dispersions of our simulated 
merger remnants, for 
which $\sigma_\star \sim 290$ km/s. In the \citetalias{Thomas2016} sample the 
SMBH masses span 
from $4.2\times 10^8 M_\odot$ to $2.1\times10^{10} M_\odot$ with an average of 
$\langle M_\bullet \rangle = 3.5 \times 10^9 M_\odot$. This is in a 
similar range
to the final SMBH masses in our merger simulations: $1.7\times10^9 M_\odot < 
M_\bullet < 1.7\times10^{10} M_\odot$. 

Even though the stellar mass and the velocity dispersion in 
our simulated galaxies lie close to the mean values of 
the \citetalias{Thomas2016} observational sample, we stress that some care 
should be taken in the interpretation of our simulated core scaling relations. 
The interpretation 
boils down to the central 
question of how local the core scouring process is. Thus, the question is 
whether the properties of the host galaxy beyond $r_\mathrm{SOI}$ strongly 
affect the core formation process or 
whether the core formation process is mainly dictated by the mass of the SMBH. 
The results in \S 
\ref{section: cusp-destruction} support the latter scenario, as the central 
core 
forms rapidly compared to the crossing time 
of the merger remnant, when the SMBH binary shrinks from $a = r_\mathrm{h}$ to 
$a = a_\mathrm{h}$. 
However, on the other hand if the large-scale properties of the host galaxies 
strongly affect the core scouring process, we should not be able to reproduce 
the observed scaling relations using only single-mass progenitor galaxy models.

\subsection{ Determining the core size } \label{section: coresersic}

There are a number of widely-used methods to determine the size of the cored region in the surface brightness profile $\mu(r)$. One popular model 
is the six-parameter Core-S\'ersic profile 
(\citealt{Graham2003, Trujillo2004}), which is defined as
\begin{equation}
 \mu(r) = \mu' \left[1+\left(\frac{r_\mathrm{b}}{r}\right)^\alpha 
\right]^{\gamma/\alpha} e^{-b\left[\left(r^\alpha + 
r_\mathrm{b}^\alpha\right)/R_\mathrm{e}^\alpha\right]^{1/(\alpha n)}}.
\label{coresersic}
\end{equation}
In the outer parts, the profile follows the ordinary three-parameter 
S\'ersic profile \citep{Sersic1963, Caon1993} defined by the S\'ersic index $n$, 
the effective radius $R_\mathrm{e}$ and the overall 
normalization $\mu'$. Towards the central regions, the core-S\'ersic profile 
breaks away from the S\'ersic profile at the break radius (i.e. the core size) $r_\mathrm{b}$ to allow for a shallow central power-law with a slope $\gamma$. The parameter $\alpha$ determines how sudden this break is. Finally, the constant $b$ is determined by requiring that the effective radius $R_\mathrm{e}$ 
encloses half of the total light of the core-S\'ersic profile. 

We perform the profile fitting using the CMPFIT Library\footnote{http://www.physics.wisc.edu/~craigm/idl/cmpfit.html}. The radial fitting range  
for all the surface brightness profiles shown in Fig. \ref{fig: surfacebrightness} is $0.01$ kpc $< r < 62$ kpc in this Section, similar to \citetalias{Thomas2016}. For all of these fits the core-S\'ersic index is fixed to $n=4$. The reason for this procedure is to avoid any possible degeneracy between the S\'ersic index 
and the break radii in the fitting procedure \citep{Dullo2012}, and to compare 
the results from subsamples $\gamma = 1$ and $\gamma = 3/2$ with each other in a 
more straightforward and robust manner. If we allow the S\'ersic index to 
freely vary, we typically get $n \leq 3$ for the $\gamma=1$ simulation sample, which is on 
the low side for massive early-type galaxies, see e.g. the Appendix A of 
\cite{Naab2006} for details. The initial surface brightness profiles of the $\gamma=3/2$ sample are best fit with the S\'ersic indices close 
to $n=4$ \citep{Dehnen1993}, making these initial conditions more in tune what would be expected for massive early-type galaxies. 

Another option is to fit the so-called Nuker profile \citep{Lauer1995} to the surface brightness profile, defined as
\begin{equation}
\mu(r) = \mu_\mathrm{b} 2^{(\beta-\gamma)/\alpha} \left( \frac{r}{r_\mathrm{b}} \right)^{-\gamma} \left[ 1+\left( \frac{r}{r_\mathrm{b}} \right)^\alpha  \right]^{(\gamma-\beta)/\alpha},
\end{equation}
where $r_\mathrm{b}$ is the break radius, $\gamma$ is the logarithmic slope of the profile inside and $\beta$ is the logarithmic slope 
outside the break radius. The parameter $\alpha$ again measures the sharpness of the transition from the inner to the outer power-law at the break radius, $r=r_\mathrm{b}$ . The Nuker profile was originally used to fit the Hubble Space Telescope observations of early-type galaxies within the central 10 arcseconds. At the distance of NGC 1600 this corresponds to $\sim 3.1$ kpc. Later, the Nuker profile has also been used to fit the surface brightness profiles in the entire radial range of observed galaxies. However, it is well known
that the parameters of the Nuker profile depend on the radial fitting range \citep{Graham2003}.

\begin{figure}
\includegraphics[width=\linewidth]{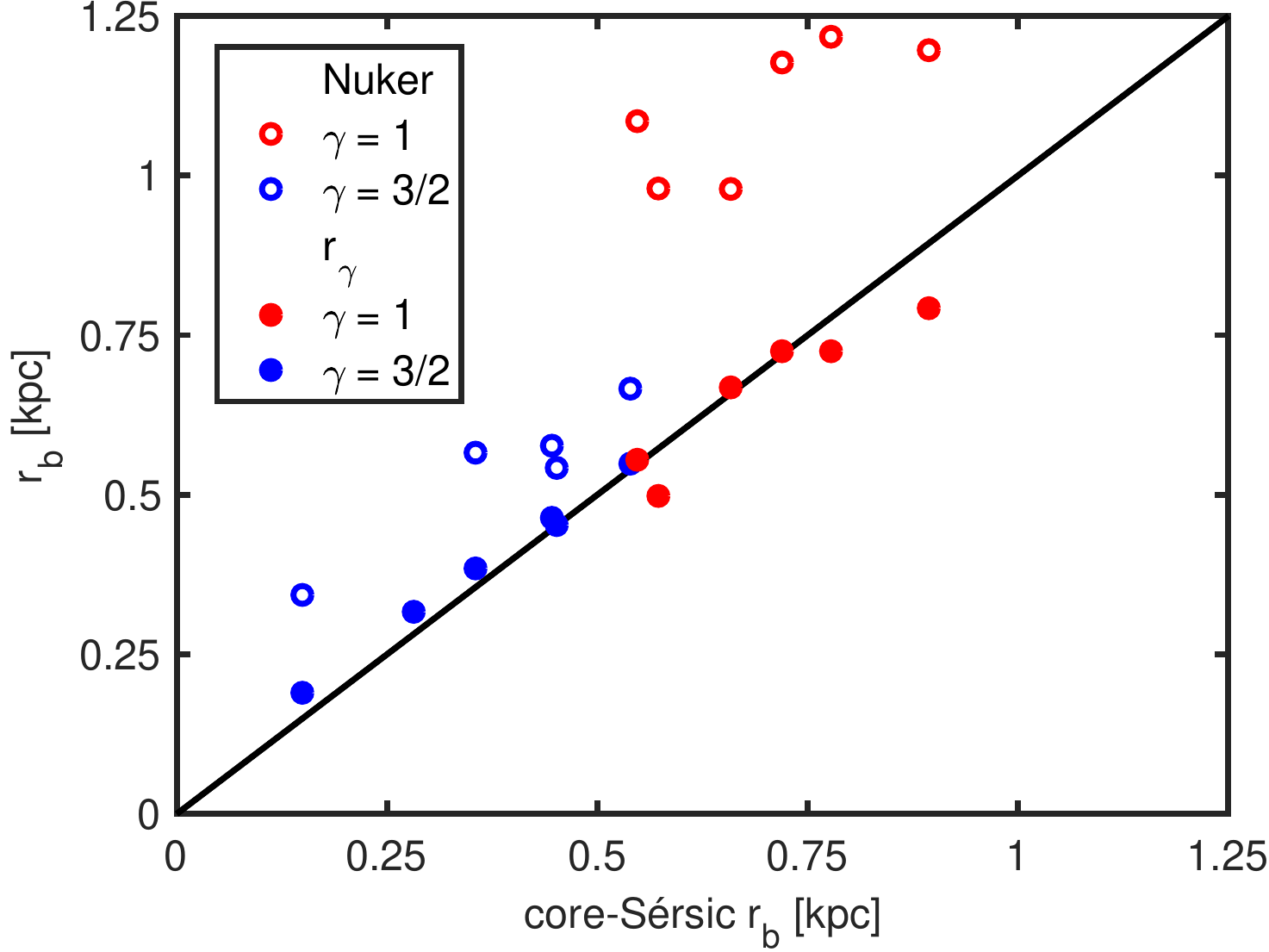}
\caption{A comparison of different core size determination methods for our simulated surface brightness profiles. The core radii obtained using the non-parametric $r_\gamma$ method agree well with the core-S\'ersic break radii $r_\mathrm{b}$, whereas the Nuker break radii are systematically larger. Note that the Nuker core sizes strongly depend on the selected radial fitting range, as discussed in the text. Here the fitting range is $0.01$ kpc $<r <3.1$ kpc for the Nuker fit.}
\label{fig: fits}
\end{figure}

We performed the Nuker profile fits first in the same radial range as for the core-S\'ersic fits ($0.01$ kpc $<r< 62$ kpc). In this case the best-fit Nuker profiles for all the galaxies in our simulated 
galaxy sample have $\alpha<1$. In general, when $\alpha \lesssim 1$, the presence of two power-laws may not emerge and instead one continuous curving arc describes the profile \citep{Graham2003}. Consequently, the break between the inner and outer profiles is not well defined. We tested 
whether stacking of the surface brightness profiles from different viewing angles or using elliptical rather than circular bins in Fig. \ref{fig: surfacebrightness} 
would affect the results, but we still found $\alpha<1$ for all the Nuker fits. However, we found that when the outer edge of the radial fitting range is chosen 
to be below $2\times R_{\mathrm{e}}$, the sharpness parameter $\alpha$ becomes larger than unity. The core size $r_\mathrm{b}$ thus also depends on the radial fitting range, which is consistent with findings in the literature (\citealt{Graham2003}).  Next, we performed the Nuker fits only in the central parts of the surface brightness profile in the range $0.01$ kpc $<r<3.1$ kpc. Now we found $\alpha>1$ for all the simulated galaxies expect $\gamma$-1.5-BH-2. The core sizes obtained from the core-S\'ersic and the central Nuker fits are compared in Fig. \ref{fig: fits}. We find that the Nuker break radii are systematically larger than the core-S\'ersic core radii, especially in the simulation set with $\gamma = 1$.

Finally, a simple non-parametric method can also be used to estimate the core size by studying the logarithmic slope 
of the surface brightness profile. The ''cusp radius'' $r_\gamma$ can then be defined as 
\begin{equation}
\left[ \derfrac{\log{\mu(r)}}{\log{r}} \right]_{r=r_\gamma} = -\frac{1}{2},
\end{equation}
i.e. as the point where the logarithmic slope of the surface brightness profile, $\mu(r)$, equals $-1/2$ (\citealt{Carollo1997, Lauer2007}). We find that the cusp radii $r_\gamma$ of our simulated galaxies are consistent with the core radii derived from the core-S\'ersic fits (see Fig. \ref{fig: fits}). 

In this study, we use the core-S\'ersic break radius $r_\mathrm{b}$ to measure the sizes of the cores of our simulated galaxies. This selection makes the comparison with \citetalias{Thomas2016} straightforward. The Nuker fits are not used in the following sections because of their dependence on the radial fitting range and the $\alpha < 1.0$ problem encountered while fitting our simulated surface brightness profiles. The cusp radii $r_\gamma$ could also be used in the following analysis, as the core sizes using the method are consistent with the results obtained using the core-S\'ersic fit, which was also found by  \citetalias{Thomas2016}.

\subsection{Simple scouring model}\label{section: simple}

We use a simple core scouring model to interpret the results of our 
simulations. Starting from the stellar density profile $\rho(r)$ in our initial 
conditions we first remove a mass deficit equaling the mass of the SMBH, i.e. 
$M_\mathrm{def} = M_\bullet$. 
This choice is motivated both by observations and N-body simulations (see e.g. 
\citealt{Merritt2013}). The stellar mass is removed from the density profile in 
such a way that the central part of the scoured density profile 
$\rho_\mathrm{s}(r)$ becomes flat up to the deficit radius $r_\mathrm{def}$:
\begin{equation}
\rho_\mathrm{s}(r) = \begin{cases}
\rho(r), & r \geq r_\mathrm{def}\\
\rho(r_\mathrm{def}), & r < r_\mathrm{def}.
\end{cases}
\label{eq: flattened-density}
\end{equation}
The mass deficit is therefore
\begin{equation}
M_\mathrm{def} = 4 \pi \int_0^{r_\mathrm{def}} \left[ \rho(r) - 
\rho(r_\mathrm{def}) \right] r^2 dr,
\end{equation}
from which the mass deficit radius can be solved numerically. The deficit 
radius 
$r_\mathrm{def}$ is sometimes also referred to as the core radius (e.g. 
\citealt{Volonteri2003}), but we reserve this label for the break radius 
$r_\mathrm{b}$ obtained from the core-S\'ersic fit (Eq. \ref{coresersic}) of 
the surface brightness profile.

Next, we measure the extent of the sphere of influence of the SMBH from the 
flattened density profile $\rho_\mathrm{s}(r)$. Instead of using the
definition of $r_\mathrm{h}$ from Eq. \eqref{eq: rinfl1}, we follow here 
\citetalias{Thomas2016} and define the sphere of influence $r_\mathrm{SOI}$ 
using  
the enclosed mass as
\begin{equation}
 M_\star(r_\mathrm{SOI}) = 4 \pi \int_0^{r_\mathrm{SOI}} \rho_\mathrm{s}(r)r^2 
dr = M_\bullet.
\label{rsoi}
\end{equation}
The sphere of influence $r_\mathrm{SOI}$ is then again most conveniently solved 
numerically.

The two-dimensional surface brightness profile $\mu(r)$ is obtained by 
projecting the analytically scoured density profile. Labeling the constant 
stellar 
mass-to-light ratio as $M_\star/L = 
\Upsilon_\star$, the surface brightness profile can be computed from
\begin{equation}
 \mu(r) = \Upsilon_\star \Sigma(r) = 2 \Upsilon_\star \int_r^\infty 
\frac{\rho_\mathrm{s}(r') r'}{\sqrt{r'^2-r^2}} dr'.
\end{equation}
Finally, the break radii $r_\mathrm{b}$ of the generated surface brightness 
profiles are obtained from the core-S\'ersic fitting procedure as described in 
the previous section.

\subsection{Results}\label{scaling-results}

\begin{figure*}
\includegraphics[width=\linewidth]{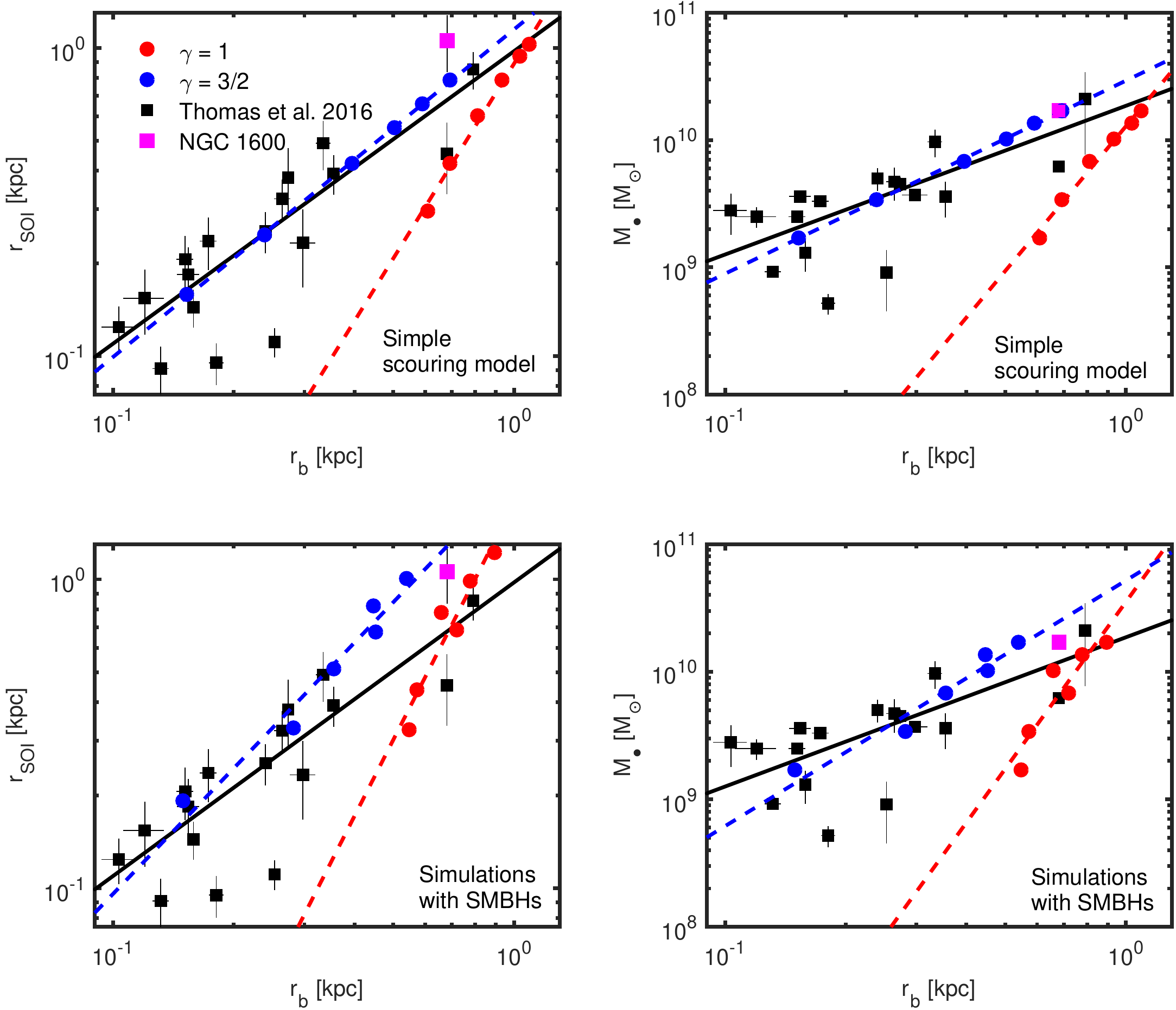}
\caption{Core scaling relations $(r_\mathrm{b}$ vs. $r_\mathrm{SOI})$ in the 
left 
column and $(r_\mathrm{b}$ vs. $M_\bullet)$ in the right column. The top panels 
present the relations obtained using the simple scouring model described in 
\S \ref{section: simple}. The bottom panels show the relations obtained 
from the 
merger simulations. The observed galaxies by \citetalias{Thomas2016} are marked 
as black squares with 1-sigma uncertainties. The fitted scaling relation is 
shown 
as the black solid line. NGC 
1600 is highlighted as the large magenta square in each plot. The relations 
from the simple scouring model agree qualitatively very well with the merger 
simulation 
relations. For steeper initial stellar density profiles $(\gamma=1.5)$ the 
agreement between both the analytic and simulated slopes is better with the 
observations.}
\label{fig: relations}
\end{figure*}

\begin{table*}
\begin{center}
\begin{tabular}{ |l c c| }
  \hline
  \multicolumn{3}{|c|}{$ \log_{10}(r_\mathrm{SOI}/\mathrm{kpc}) = a_1+b_1 
\log_{10}(r_\mathrm{b}/\mathrm{kpc})$} \\
  \hline
   & $a_1$ & $b_1$\\
  \cite{Thomas2016} & $-0.01 \pm 0.29$& $0.95\pm0.08$\\
  Simple model $\gamma=1$ & $-0.05 \pm 0.01$& $2.10\pm0.08$\\
  Simulation $\gamma=1$ & $0.25 \pm 0.03$& $2.55\pm0.22$\\
  Simple model $\gamma=3/2$ & $0.06 \pm 0.01$& $1.06\pm0.01$\\
  Simulation $\gamma=3/2$ & $0.33 \pm 0.07$& $1.35\pm0.17$\\  
  \hline
\end{tabular}
\begin{tabular}{ |l c c| }
  \hline
  \multicolumn{3}{|c|}{$ \log_{10}(M_\bullet/M_\odot) = a_2+b_2 
\log_{10}(r_\mathrm{b}/\mathrm{kpc})$} \\
  \hline
   & $a_2$ & $b_2$\\
  \cite{Thomas2016} & $10.27 \pm 0.51$& $1.17\pm0.14$\\
  Simple model $\gamma=1$ & $10.11 \pm 0.01$& $3.78 \pm0.19$\\
  Simulation $\gamma=1$ & $10.55 \pm 0.09$& $4.35\pm0.58$\\
  Simple model $\gamma=3/2$ & $10.47 \pm 0.01$& $1.52\pm0.01$\\
  Simulation $\gamma=3/2$ & $10.72 \pm 0.11$& $1.93\pm0.28$\\  
  \hline
\end{tabular}
\caption{The parameter values of the core scaling relations 
($r_\mathrm{b}$ vs. $r_\mathrm{SOI}$) and 
($r_\mathrm{b}$ vs. $M_\bullet$) presented in Fig. \ref{fig: relations}. The 
constant 
term $a$ from both the simple scouring model and the merger simulations are 
within the uncertainty of the observed relation of \citetalias{Thomas2016}. 
We also see a trend for both the analytic and simulated slopes $b$ towards the 
observed slope values of $b\sim 1$ for steeper initial stellar density 
profiles, 
with the
steeper $\gamma=1.5$ profiles providing clearly a better match with the 
observations.} 
\label{table: relations}
\end{center}
\end{table*}

We derive the break radii $r_\mathrm{b}$ of the surface brightness profiles
of our simulated merger remnants using the fitting procedure described in 
\S \ref{section: coresersic}. The residuals of the fits are of the same 
order as the residuals of the observed core-S\'ersic of NGC 1600 presented in 
\citetalias{Thomas2016}, that is below $\Delta \mu < 0.1$ mag. The extents of 
the spheres of 
influence $r_\mathrm{SOI}$ are obtained directly from the simulation data using
the enclosed stellar mass definition (see Eq. \ref{rsoi}). 
These parameter values comprise the core data points 
($r_\mathrm{b},r_\mathrm{SOI}$), 
($r_\mathrm{b},M_\bullet$) of the merger simulations. 

We produce another set of ($r_\mathrm{b},r_\mathrm{SOI}$), 
($r_\mathrm{b},M_\bullet$) data points using 
our simple scouring model and the 
12 initial conditions of the merger simulations containing central SMBHs. The 
actual scaling relations for both the simulation samples and the analytical 
scouring models
are obtained by fitting the free parameters $a_1,b_1,a_2,b_2$ in the fitting 
functions
\begin{equation}
 \begin{aligned}
  \log_{10}(r_\mathrm{SOI}/\mathrm{kpc}) = a_1 + b_1 
\log_{10}(r_\mathrm{b}/\mathrm{kpc}) \\
\log_{10}(M_\bullet/M_\odot) = a_2 + b_2 
\log_{10}(r_\mathrm{b}/\mathrm{kpc}).
 \end{aligned}
\end{equation}
The values of the best fit parameters are listed in Table \ref{table: 
relations}, where 
the error estimates have been computed using a simple bootstrap method. The 
results from both the merger runs and the simple scouring model are 
presented in Fig. \ref{fig: relations}. 
The simulation data points and the simple scouring 
model points are plotted as red ($\gamma=1$) and blue ($\gamma=1.5$) filled 
circles, respectively. 
The scaling relations for both of the $\gamma=1$ 
and $\gamma=3/2$ subsamples are plotted as dotted lines of the corresponding 
color. 
The observed galaxy sample of \citetalias{Thomas2016} is shown as black 
squares, 
with the solid lines 
representing the observed scaling relations 
$\log_{10}(r_\mathrm{SOI}/\mathrm{kpc}) = 
(-0.01\pm0.29) + (0.95\pm0.08)\log_{10}(r_\mathrm{b}/\mathrm{kpc})$ and 
$\log_{10}(M_\bullet/M_\odot) = (10.27\pm0.51) + 
(1.17\pm0.14)\log_{10}(r_\mathrm{b}/\mathrm{kpc})$.

Studying Fig. \ref{fig: relations}, we see that our simple scouring model 
produces qualitatively similar results to the major merger simulations. 
We also see a trend towards the observed relations with increasing initial 
stellar 
density profile slope $\gamma$, as the scaling relations flatten for steeper 
initial stellar profiles.
This fact points towards very cuspy initial stellar 
density profiles of the progenitor galaxies, with $\gamma$ in excess of $3/2$, 
in the framework of a single generation of galaxy mergers. Mergers of galaxies 
with relatively flat Hernquist-like 
stellar bulges ($\gamma=1$) result in too steep core scaling relations. Finally, we also note that our core scaling relations are consistent with the results of \cite{Lauer2007}, who obtained the $M_\bullet$-$r_\gamma$ -relation from a sample of 11 observed core galaxies using directly determined SMBH masses.

\subsection{Central homology of power-law ellipticals}

The surprisingly good agreement between the scaling relations derived from
our cuspier $(\gamma=1.5)$ initial conditions and the observed relations 
requires an explanation. All the progenitor galaxies in our merger simulation 
sample 
had identical stellar density profiles and in practise identical velocity 
profiles
outside of $r>r_\mathrm{SOI}$, while the observed \citetalias{Thomas2016} 
galaxy 
sample had 
a broad range of stellar mass and velocity dispersions, as described in \S 
\ref{section: 
scaling-samples}. 

Both observations and simulations have shown that the total density profiles of 
power-law galaxies are well represented with simple density profiles $\rho 
\propto r^{-\gamma}$ over a 
wide range in radius, with $\gamma \sim 2.0$-$2.2$ (e.g. 
\citealt{Cappellari2015, Remus2013, Remus2017} 
and references therein). At $r \ll r_{1/2}$, the stellar component dominates 
the total density profile, so the stellar density profile has a central slope 
$\gamma \sim 2$. In order to estimate the mean central stellar densities $\rho$ 
of the power-law galaxies, we model the stellar component using the Dehnen 
profile as in \S \ref{section: ic} with $\gamma=2$. We sample a large number of 
progenitor galaxy models with parameters $M_\bullet$, $M_\star$ and $r_{1/2}$ 
and their uncertainties from \citetalias{Thomas2016} and \cite{Saglia2016}. 
Next, we compute the mean central $(r < 1\ \rm pc)$ 
stellar densities for these idealized galaxy models. Although the stellar mass 
increases by a factor of $40$ from $\min(M_\star^{\mathrm{T16}})$ to 
$\max(M_\star^\mathrm{T16})$, the central stellar density increases 
simultaneously only by a factor of $6.2\pm3.8$ from the most diffuse galaxy to
the galaxy with the highest central density. Thus, the central stellar density 
increases slowly 
compared to the total stellar mass. 
The population of power-law galaxies is very homogeneous in their central 
stellar density properties. At the same time, the mass of the SMBH, $M_\bullet$ 
in the 
\citetalias{Thomas2016} sample increases much more significantly, by a factor 
of 
$\sim 50$.

We argue that this homology of the central properties of the power-law 
ellipticals is responsible for the good correspondence between the core scaling 
relations from our single-$M_\star$ merger simulation sample and the scaling 
relations 
of the observed core galaxy population. Our results support the scenario in 
which the 
progenitors of the most massive core ellipticals are cuspy power-law galaxies 
($\gamma>3/2$) with central SMBHs. The central properties of the stellar 
population are fairly homogeneous over a large range in $M_\star$, while the 
SMBH masses steadily increase as $M_\star$ grows. Thus, the SMBH mass 
$M_\bullet$ 
stands out as the main parameter in the core scouring process and is 
also responsible for determining the core scaling relations.

\section{Conclusions}\label{section: conclusions}

We have performed a series of galaxy merger simulations that include 
supermassive black holes using the regularized tree code \ketju. 
The study presented in this paper mainly concentrated on two parameters: 
the initial SMBH mass $M_\bullet$ and the central slope of the initial stellar 
density profile, $\gamma$. The properties of the progenitor galaxies were 
selected 
in such a way that the simulated merger remnant would be a close analogue to the
observed core elliptical galaxy NGC 1600, which was extensively studied in 
\citetalias{Thomas2016}.

From the simulated merger remnants we find a systematic decrease in the
central surface brightness and increasingly more tangentially biased central 
velocity 
anisotropies (decreasing $\beta$) as a function of increasing SMBH mass in the 
initial conditions. 
Similar dynamical trends can also be seen in the mock stellar 2D kinematic maps.
In addition, the kinematic maps also reveal decoupled, kpc-scale rotating 
regions which become more 
apparent for increasing SMBH masses. Finally, there are indications of even 
more complex kinematic subsystems in the $\gamma=3/2$ merger remnants with the 
most massive
initial SMBH masses. Overall, we find the best match between our 
simulated merger remnants and the observed NGC 1600 galaxy for the simulation 
run with the 
SMBH masses inferred for NGC 1600 by the dynamical modeling of 
\citetalias{Thomas2016}. 
For a single generation of mergers, especially the velocity anisotropy profiles 
favor cuspy initial stellar 
profiles 
for the progenitor galaxies with very steep profiles in excess of $\gamma 
\gtrsim 3/2$.

We also study the time evolution of the core scouring process in detail. Our 
simulated galaxy mergers with 
larger initial SMBH masses result is merger remnants with systematically larger 
cores, in good agreement with 
earlier findings (e.g. \citealt{Merritt2006}). Defining the core formation as 
the moment when the central density profile flattens, 
the formation process is very rapid and 
occurs when the semi-major axis of SMBH binary shrinks from the influence 
radius 
to the hard radius, 
i.e. from $a=r_\mathrm{h}$ to $a=a_\mathrm{h}$. When the SMBH binaries become 
hard, the destruction of the radial 
orbits commences and the core regions of the merger remnants start to develop 
a tangentially biased stellar orbit population. The development of the 
anisotropic velocity profile is a slow process 
compared to core formation, and occurs mostly after the central density profile 
has already become flat.

We also inspect the core scaling relations ($r_\mathrm{b}, r_\mathrm{SOI}$) and 
($r_\mathrm{b}, 
M_\bullet$) between the core size, the SMBH mass and its sphere of influence. 
We find similar results both for actual merger simulations and a simple 
analytical scouring model.
Our simulated relations become increasingly more shallow for steepening initial 
stellar density profile slope, with
the $\gamma=1.5$ runs being in better agreement with the observations than the 
shallower $\gamma=1$ sample, when $N_{\mathrm{mergers}}=1$ merger generations 
is 
assumed. 
This is in support of the scenario in which power-law ellipticals with steep 
$\gamma \sim 2$ profiles are the progenitors of the core galaxy population.

The agreement between our simulated core scaling relations and the observed 
relations are very good, especially
considering that the observed galaxy sample of \cite{Thomas2016} consisted of 
core 
ellipticals in a wide stellar-mass range, whereas all of our simulated merger 
remnants had 
stellar masses comparable to NGC 1600. This could be related to the homology of 
the central properties of power-law ellipticals, for which the central stellar 
densities increase only modestly for significant increases in the stellar and 
central 
black hole masses. This also
demonstrates that core formation is a very local process, for which the most 
important quantity is the mass of the central supermassive black hole. 

The core scaling relations and the scouring scenario that relies on the central 
homology of power-law ellipticals will be inspected 
further in a follow-up study, which contains a broader and more realistic 
sample 
of merger progenitor galaxies. 
Selecting the initial SMBH 
masses from the $M_\bullet$ - host galaxy scaling relations, this future study 
will focus on varying 
also the stellar masses $M_\bullet$ and dark matter fractions $f_\mathrm{DM}$ 
inside the effective radii of the progenitor galaxies.
 We also plan to simulate re-mergers of already scoured merger remnants, as 
realistic 
massive galaxies are expected to have experienced a large number of mergers 
since their initial assembly.

As already discussed by the pioneering study of \cite{Milosavljevic2001}, 
high-resolution mergers of power-law galaxies 
with $\gamma=2$ that include accurate dynamics inside the sphere of influence 
of 
the SMBHs is a notoriously difficult problem 
to simulate. Our results favor steep central stellar profile slopes 
$\gamma>3/2$ 
for the progenitors of massive core ellipticals. Constructed with the same 
parameters and mass resolution as the $\gamma=3/2$ NGC 1600 initial conditions, 
the $\gamma=2$ progenitor 
galaxies would contain up to a factor of $\sim 14$ more stellar particles in 
the 
regularized region. The resulting increase of
a factor of $\sim 200$-$300$ for the computational load  in the \archain{}
calculation \citepalias{Rantala2017} is unfeasible with current simulation 
codes.
Pursuit towards simulating steep $\gamma=2$ profiles poses a numerical 
challenge, which motivates further 
development of simulation codes.

\small
\begin{acknowledgements}
The authors thank the anonymous referee for constructive comments on the manuscript.
The numerical simulations were performed on facilities hosted by 
the CSC -IT Center for Science in Espoo, Finland. A. R. acknowledges support 
from the MPA Garching Visitor Programme. A. R. is funded by the doctoral 
programme of Particle Physics and Universe Sciences at the University of 
Helsinki. P. H. J. and A. R. acknowledge the support of the Academy of Finland 
grant 
274931. T. N. acknowledges support from the DFG Cluster of Excellence 'Origin 
and Structure of the Universe'.
\end{acknowledgements}

\normalsize

\bibliography{references}

\end{document}